\RequirePackage{amsmath} 
\documentclass[envcountsame]{llncs} 
\usepackage{algorithm,amsfonts,amsmath,amssymb,bm,verbatim}
\usepackage{cellspace,chngcntr,colortbl,epsfig,graphicx,mathtools,xr-hyper,hyperref,multirow,tabularx,tikz,xcolor}
\usepackage{arydshln}
\usepackage[misc]{ifsym}
\usepackage[all]{xy}
\usepackage[noend]{algpseudocode}

\usepackage{makecell}

\renewcommand{\v}[1]{{\bm #1}}
\renewcommand{\qed}{\hfill\square}
\newcommand{\keywords}[1]{\par\addvspace\baselineskip\noindent\keywordname\enspace\ignorespaces#1}
\spnewtheorem{assumption}[theorem]{Assumption}{\bfseries}{\itshape}

\DeclarePairedDelimiter\ket{\lvert}{\rangle}
\DeclarePairedDelimiterX\braket[2]{\langle}{\rangle}{#1 \delimsize\vert #2}

\newcommand{\nn}{\nonumber \\}
\newcommand{\algrule}{\Statex\par\vskip2pt\hrule\par\vskip-2pt}
\algrenewcommand\algorithmicrequire{\textbf{Input:}}
\algrenewcommand\algorithmicensure{\textbf{Output:}}

\graphicspath{{./}{../figures/}{./figures/}{../../figures/}} 

\begin{document}

\title{A Quantum Circuit for Gaussian Elimination}

\author{Hochang Lee \and Kyung Chul Jeong \and Panjin Kim$^{(\text{\Letter})}$}
\institute{The Affiliated Institute of ETRI, Daejeon 34044, Korea \\
pansics@nsr.re.kr}

\maketitle
\begin{abstract}
A quantum circuit for Gaussian elimination is developed in this work.
While previous quantum implementations of Gaussian elimination were restricted to $\mathrm{GF}(2)$, the proposed design is generalized to any finite field.
Compared with previous works in $\mathrm{GF}(2)$, the developed circuit achieves garbage-free construction while retaining the best known asymptotic Toffoli depth up to logarithmic factors.
%
\end{abstract}

\section{Introduction}\label{sec:1}
While an abstract quantum algorithm is described by a series of unitary operations, its implementation often involves a classical reversible construction of $x$-bit input $y$-bit output map $f:\{0,1\}^x \rightarrow \{0,1\}^y$.
For example, there exists a large number of cryptanalysis results in which the quantum implementation of vectorial Boolean functions is the central theme\;\cite{aes16,sha16,sch16,ecdlp17,bern19,jaq20}.
A study of such low-level implementations helps us accurately estimate the performance of the algorithm\;\cite{gid21}, or even in some cases gives the community a chance to reassess the soundness of the model or the assumption that the algorithm is based on\;\cite{bht,cns}.

Assume a field element is represented by a bit string and a matrix is encoded by an even longer bit string.
Gaussian elimination can then be viewed as a map $f:\{0,1\}^x \rightarrow \{0,1\}^y$, where $y$ is strictly smaller than $x$ since the map is not injective.
The size of work space for Gaussian elimination to be reversible is thus at least $x-y$.
Previous works on quantum circuits for Gaussian elimination (which only exist for $\text{GF}(2)$) are either unable to meet the minimum work (garbage) space or to achieve the same complexity as the classical implementation.

This paper investigates the quantum implementation of Gaussian elimination in \emph{any finite field}.
The results are summarized as follows:
\begin{itemize}
  \item[$\bullet$] A pseudo row echelon form is introduced to turn the inherently non-injective Gaussian-elimination mapping into a one-to-one transformation while preserving the row-echelon information required for solving linear systems. A reversible construction for the pseudo row echelon form is developed.
  \item[$\bullet$] No garbage space is occupied other than the input qubits which means the result of Gaussian elimination is overwritten on the input, and the information that required to reverse the operation is held by the minimum number of qubits.
  \item[$\bullet$] In $\mathrm{GF}(2)$, the developed circuit turns out to be advantageous over previous works either in Toffoli depth or number of qubits.
\end{itemize}

\begin{table}[htbp]
	\label{tab:complexity-comparison}
	\caption{Circuit complexities for reversible Gaussian elimination of a matrix in $\text{GF}(2)^{m\times n}$, $m\ge n$. Temporary space row reads the number of work qubits that are zeroed before and after Gaussian elimination.
		In\;\cite{GE21}, multi-controlled NOT gate is adopted which is further decomposed into Toffoli gates with (\cite{GE21}) and without (\cite{Claudon2024}) temporary space.
		Supplemental (garbage) space row reads the number of work qubits that are initially zero but not returned to zero at the end of Gaussian elimination.
		The classical arithmetic complexity of Gaussian elimination is $O(n^2 m)$.}
	\centering
	\renewcommand{\arraystretch}{1.3}
	\begin{tabular}{
			m{0.4cm}
			>{\centering}m{1.35cm} |
			>{\centering}m{2.4cm}
			>{\centering}m{3.4cm}
			>{\centering}m{1.7cm}
			>{\centering\arraybackslash}m{1.7cm} }
		&& \scriptsize{\makecell{Toffoli count}}
		& \scriptsize{\makecell{Toffoli depth}}
		& \scriptsize{\makecell{Temporary\\space}}
		& \scriptsize{\makecell{Supplemental\\space}}\\\hline
		\cite{GE21} & \tiny{\makecell{With\\ temp. space}} &
		$O(n^{2}m^{2})$ & $O(n^{2}m \log_{2}\!m)$ & $m-2$ & $0$
		\\
		\cite{GE21} & \tiny{\makecell{Without\\ temp. space}} &
		$O(n^{2}m^{2}(\log_{2}\!m)^{2})$ & $O(n^{2}m (\log_{2}\!m)^{3})$ & $0$ & $0$
		\\
		\multicolumn{2}{c|}{\cite{GE23}} &
		$O(n^{2}m)$ & $O(nm)$ & $0$ & $\tfrac{n(2m-n-1)}{2}$
		\\
		\multicolumn{2}{c|}{\cite{GE22_BJ21}} &
		$O(n^{2}m)$ & $O((m+n)\log_{2}\!n)$ & $m+\tfrac{n(n+1)}{2}$ & $m+\tfrac{n(n+1)}{2}$
		\\
		\multicolumn{2}{c|}{This work} &
		$O(n^{2}m)$ & $O(m\! \log_2\!n + n (\log_{2}\!m)^4)$ & $m-1$ & $0$
		\\
	\end{tabular}
\end{table}

A quantitative summary is given by Table\;\ref{tab:complexity-comparison}.
The paper is organized as follows.
Section\;\ref{sec:2} covers the necessary notation and definition as well as cost measure and previous works.
Section\;\ref{sec:3} introduces an algorithm for row echelon form, followed by the reversible constructions in Section\;\ref{sec:4}.
We conclude the paper with some discussions in Section\;\ref{sec:5}.

\section{Preliminaries}\label{sec:2}
\subsection{Gaussian Elimination}
Gaussian elimination is an algorithm for solving a system of linear equations, which has various usages for computing rank, determinant or inverse of a matrix.
Consider the following row operations:
\begin{itemize}
  \item Swapping two rows
  \item Multiplying a row by a field element
  \item Adding a row by another row
\end{itemize}
Gaussian elimination is a series of the row operations to achieve the row echelon form of the matrix defined as follows~\cite{Leon}:
\begin{enumerate}
  \item The first nonzero entry in each nonzero row is 1.
  \item The number of leading zero entries in row $(k+1)$ is greater than the number of leading zero entries in row $k$.
  \item If there are rows whose entries are all zero, they are below the rows having nonzero entries.
\end{enumerate}
The algorithm is straightforward, thus its description is omitted~\cite{Leon}.
Counting the number of arithmetic operations, its complexity is $O(n^2 m)$, where $m, n$ are number of rows and columns, respectively.

In building a quantum circuit for Gaussian elimination, one needs to be aware that some row operations take place conditionally.
For example, whether or not an addition by another row is performed is dependent on the leading entry of the two rows.
In classical implementation it can simply be handled by \emph{if} clauses, but in quantum case evaluating and performing such conditional operations are not immediately conceivable.
In Section\;\ref{sec:3}, a quantum circuit for Gaussian elimination is designed, circumventing the issues relevant to the conditional operations while retaining the same complexity as in classical implementations.

\subsection{Notation and Definition}
Elementary quantum gates are used throughout the paper, but not covered here.
Unfamiliar readers may find the book by Nielsen and Chuang useful~\cite{bookchuang}.
We will work in a finite field $\text{GF}(p^l)$, for prime $p$ and $l\in \mathbb{Z}^+$.

Let $a_{ij}$ be the $i$th row, $j$th column element of $\mathbf{A} \in \text{GF}(p^l)^{m \times n}$.
We use a notation $\mathbf{A}_{[i \,;\, j ]}$ to denote a submatrix of $\mathbf{A}$ that is formed from the $i$th row to the last row and from the $j$th column to the last column of $\mathbf A$.

\begin{definition}[Pseudo row echelon form]
	A matrix $\mathbf{A}'$ is in pseudo row echelon form of a full-rank matrix\footnote{Handling a rank deficient matrix is doable, but substantially complicates the construction.} $\mathbf{A}$ if the upper triangular elements of $\mathbf{A}'$ are that of the row echelon form of $\mathbf{A}$.
	\label{def:PseudoRowEchelon}
\end{definition}

\begin{table}[htbp]
	\caption{Row echelon form and pseudo row echelon form of two different input matrices in $\text{GF}(5)^{3\times 3}$.
For some inputs, Gaussian elimination results in the same matrix.
The third column is obtained by applying Algorithm\;\ref{alg:main-algorithm} (see, Section\;\ref{sec:3}). }
	\centering
	\begin{tabular}{>{\centering}p{2cm} | >{\centering}p{2.8cm} | >{\centering}p{3.8cm}}
             & \small{Row echelon form} & \small{Pseudo row echelon form}
		\tabularnewline\hline
        $\left(\begin{matrix}
			0 & 3 & 1 \\ 2 & 1 & 0 \\ 2 & 4 & 3
		\end{matrix}\right)$
		&
        $\left(\begin{matrix}
			1 & 3 & 0 \\ 0 & 1 & 2 \\ 0 & 0 & 1
		\end{matrix}\right)$
		&
        $\left(\begin{matrix}
			\colorbox[rgb]{0.7,0.7,0.7}{\makebox(0,6){0}} & 3 & 0 \\
			\colorbox[rgb]{0.7,0.7,0.7}{\makebox(0,6){2}} & \colorbox[rgb]{0.7,0.7,0.7}{\makebox(0,6){3}} & 2 \\
			\colorbox[rgb]{0.7,0.7,0.7}{\makebox(0,6){2}} & \colorbox[rgb]{0.7,0.7,0.7}{\makebox(0,6){4}} & \colorbox[rgb]{0.7,0.7,0.7}{\makebox(0,6){4}}
		\end{matrix}\right)$
		\tabularnewline\hline
		$\left(\begin{matrix}
			1 & 3 & 0 \\ 2 & 3 & 4 \\ 3 & 3 & 0
		\end{matrix}\right)$
		&
		$\left(\begin{matrix}
			1 & 3 & 0 \\ 0 & 1 & 2 \\ 0 & 0 & 1
		\end{matrix}\right)$
		&
		$\left(\begin{matrix}
			\colorbox[rgb]{0.7,0.7,0.7}{\makebox(0,6){1}} & 3 & 0 \\
			\colorbox[rgb]{0.7,0.7,0.7}{\makebox(0,6){2}} & \colorbox[rgb]{0.7,0.7,0.7}{\makebox(0,6){1}} & 2 \\
			\colorbox[rgb]{0.7,0.7,0.7}{\makebox(0,6){3}} & \colorbox[rgb]{0.7,0.7,0.7}{\makebox(0,6){3}} & \colorbox[rgb]{0.7,0.7,0.7}{\makebox(0,6){3}}
		\end{matrix}\right)$
		\tabularnewline
	\end{tabular}
\label{tab:pREF}
\end{table}

The reason for introducing Definition\;\ref{def:PseudoRowEchelon} is that, Gaussian elimination is not injective as can be seen from Table\;\ref{tab:pREF}. 
Definition\;\ref{def:PseudoRowEchelon} does not specify the lower triangular and diagonal parts (thus it is not unique), and it will soon be shown in Section\;\ref{sec:3} that some information is saved in these parts so that the whole process is reversible.
Note that pseudo row echelon form can be used to solve the system of linear equations as equally good as row echelon form.
%

%
\begin{definition}[Pivot index]
\label{def:pivotid}
  Consider the first column vector of $\mathbf{A}$, i.e., \\ $(a_{11} , a_{21}, \ldots, a_{m1})$.
  Pivot index of $\mathbf A$ is the smallest $i$ such that $a_{i1} \ne 0$.
\end{definition}
Additionally, we define the division symbol in finite fields such that it works for additive identity.
\begin{align}
  /:\text{GF}(p^l) \times \text{GF}(p^l) \rightarrow \text{GF}(p^l),
~~~~
  a/b =
  \left\{
    \begin{array}{ll}
      a,        & \hbox{if $b=0$;} \\
      a b^{-1}, & \hbox{otherwise.}
    \end{array}
  \right.
\label{eq:division}
\end{align}

\subsection{Cost Measure}
A universal gate set we adopt is \{NOT, CNOT, Toffoli\}~\cite{NCT}.
Multi-controlled NOT gate with $m$ control qubits (denoted by $C^m\!X$) will also be used, but that will eventually be decomposed into NOT, CNOT, Toffoli gates. 

Among the gates, only Toffoli gate will account for the cost in this work as the logically robust Toffoli gate is expected to be costly to implement~\cite{JVF12}.
It is possible to further decompose a Toffoli gate into smaller gates such as Hadamard, T, and CNOT gates~\cite{bookchuang}, optionally with extra work qubits~\cite{AMMR13,Sel13,AMM14}.
Developing and optimizing such low level circuits is an active area of research~\cite{nam18,GMM22} which is beyond the scope of this work.

In analyzing time complexity, the number of non-parallelizable operations known as depth is considered.
We work in rather high level fashion and analyze the complexity using operation depth (such as multiplicative depth) when possible.
Space complexity is simply accounted for by the number of qubits.

\subsection{Previous works}
\label{sec:2.4}
A quantum implementation of Gaussian elimination has been studied only in $\text{GF}(2)$.
In 2021, Esser et al. came up with a quantum circuit for Gaussian elimination in order to develop a quantum algorithm for syndrome decoding problem~\cite{GE21}, which was further improved by Perriello et al. in 2023~\cite{GE23}. 
The main idea is to implement the information set decoding algorithm~\cite{Prange} with the classical exhaustive search part being replaced by Grover search~\cite{Grover}.
In 2022, a separate line of research focused on symmetric-key primitives. Bonnetain and Jaques designed a Gaussian-elimination circuit over $\text{GF}(2)$ to serve as the linear-algebraic subroutine of the offline Simon algorithm~\cite{GE22_BJ21}.
In this setting, it is sufficient to determine whether the vectors sampled by Simon’s algorithm span the entire vector space.

The circuit developed in this work applies to any finite fields.
However, since the quantum circuit for Gaussian elimination has only been studied in $\text{GF}(2)$, comparisons are given in $\text{GF}(2)$.
In Table\;\ref{tab:complexity-comparison}, the circuit complexities are briefly summarized.

Note that in~\cite{GE23}, the authors dealt with square matrices (i.e., $m=n$) and used $n(n-1)/2$ work qubits which take up the supplemental space.
By observation, their circuit can be modified to use at most $m-1$ temporary work qubits retaining the same asymptotic gate and depth complexities, by restoring index qubits after each row reduction.

\section{Algorithm}\label{sec:3}

An algorithm for pseudo row echelon form is given in this section.
Here we focus on specifications of subroutines, and postpone developing the quantum implementation of each one in the next section.
The following algorithm takes as input a matrix $\mathbf{A} \in \text{GF}(p^{l})^{m \times n}$, where $l,m,n\in \mathbb{Z}^+$ and $p$ is a prime number, and outputs its pseudo row echelon form.

%
\begin{algorithm}[H]
	\small
	\caption{$\mathsf{PseudoRowEchelonForm}$}
	\begin{algorithmic}[1]
		\Require $\mathbf{A} \in \text{GF}(p^{l})^{m \times n}$
		\Ensure $\mathbf{A}'' \in \text{GF}(p^{l})^{m \times n}$
		\algrule
		\State $\mathbf{A}' \gets \mathbf{A}$, $\mathbf{A}'' \gets \{0\}^{m\times n}$
		\For {$j \in \{1, \ldots, n\}$}
		\Comment{column iterator}
		\State $\mathmakebox[0pt][l]{u_j}\phantom{\mathbf{A}'} \gets \hyperref[alg:labeling]{\mathsf{Labeling}}(\mathbf{A}')$
		\State $\mathbf{A}' \gets \hyperref[alg:pivoting]{\mathsf{Pivoting}}(\mathbf{A}', u_{j})$
		\State $\mathbf{A}' \gets \hyperref[alg:row reduction]{\mathsf{RowReduction}}(\mathbf{A}')$
        \State delete $u_j$
		\State $\mathbf{A}''_{[j\,;\, j]} \gets \mathbf{A}'$
		\vspace{1mm}
		\State $\mathbf{A}' \gets \mathbf{A}'_{[2\, ;\, 2]}$
		\Comment{ignored if $j=n$}
		\EndFor
		\State \Return $\mathbf{A}''$
	\end{algorithmic}
	\label{alg:main-algorithm}
\end{algorithm}
\noindent
The correctness proof showing that the output of Algorithm~\ref{alg:main-algorithm} satisfies Definition~\ref{def:PseudoRowEchelon} is provided in Appendix~\ref{app:CorrectnessOfMainAlgo}.
\noindent Brief description of each subroutine is given first, followed by a proposition giving an upper bound on the quantum cost for implementing Algorithm\;\ref{alg:main-algorithm}.
Pseudo code for each subroutine can be found in Appendix\;\ref{sec:app-ImplDetail}.


\paragraph{\textbf{Labeling}}
Given an input matrix $\mathbf{A}$, $\mathsf{Labeling}$ finds the pivot index (Definition\;\ref{def:pivotid}) of $\mathbf{A}$ and save it as $u\in \mathbb{Z}^+$ in unary format.

\paragraph{\textbf{Pivoting}}
Given an input matrix $\mathbf{A}$ and an integer $u$, the goal of $\mathsf{Pivoting}$ is to swap the first and the $u$th rows of $\mathbf{A}$.

\paragraph{\textbf{Row reduction}}
Given an input matrix $\mathbf{A}$, $\mathsf{RowReduction}$ first divides $a_{i2}, a_{i3}, \ldots, a_{in}$ ($i$th row of $\mathbf{A}$ except $a_{i1}$) by $a_{i1}$ (see, Eq.\;(\ref{eq:division})), for all $i\in \{1,\ldots, m\}$.
Let the resulting matrix be $\mathbf{A}'$ and its elements denoted by $a_{ij}'$.

Once $\mathbf{A}'$ is obtained, $a_{12}',a_{13}',\ldots,a_{1n}'$ are subtracted from $a_{i2}',a_{i3}',\ldots,a_{in}'$, respectively if $a_{i1}' \ne 0$, for all $i\in \{2,\ldots, m\}$.
This is basically subtracting the first row from the $i$th row except the first column.

\paragraph{\textbf{Deleting the pivot index}}
The pivot index is no longer useful once the row reduction is completed, yet it occupies $O(m)$ temporary space.
If the pivot indices are to be saved, they will eventually occupy $O(mn)$ space as Gaussian elimination iteratively reduces a matrix $O(n)$ times.
In a quantum circuit, deleting the pivot index is in fact reversing what have been done on the first column of the matrix.
Notice that it does not involve fully reversing $\mathsf{Pivoting}$ as we only need to undo the \emph{first column}.

\bigskip
Main body of this paper concerns the reversible implementation of each line in Algorithm\;\ref{alg:main-algorithm}.
The total cost is bounded by the following proposition:
%


\begin{proposition}
	For a matrix $\mathbf{A} \in \mathrm{GF}(p^{l})^{m \times n}$, $m\ge n$, Algorithm\;\ref{alg:main-algorithm} is reversibly implementable in which the quantum cost is given by Table\;\ref{tab:complexity_par}.
	\begin{table}
		\renewcommand{\arraystretch}{1.3}
		\caption{Quantum cost for implementing Algorithm\;\ref{alg:main-algorithm}.
			Here, $\alpha = l \cdot \lceil\log_{2}p\rceil$. }
		\centering
		\begin{tabular}{
				>{\centering}m{2.0cm} |
				>{\centering}m{4.9cm} |
				>{\centering}m{4.1cm}  }
			& Count & Depth \tabularnewline\hline
			Toffoli & $O\big( n m (\log_{2}^{3} m + \alpha\log_{2}^{2}\alpha + \alpha n ) \big)$ & $O\big( n\log_{2}^{4}m + \alpha n \log_{2}m  \big)$\tabularnewline
			Multiplication & $O(n^{2} m)$ & $O(n\log_{2}n)$ \tabularnewline
			\scriptsize{Multiplicative inverse} & $O(n m)$ & $O(n)$\tabularnewline
			Addition & $O(n^{2} m)$ & $O(m\log_{2}n)$ \tabularnewline
			\scriptsize{Additive inverse} & $O(n^{2} m)$ & $O(n\log_{2}n)$
		\end{tabular}
		\renewcommand{\arraystretch}{1.0}
		\label{tab:complexity_par}
	\end{table}
	\label{prop:main_par}
\end{proposition}

Proof is given in Section\;\ref{sec:complexity}.
More detailed (non-asymptotic) bounds are given in Appendix~\ref{app:app-upper-bound}.
The condition $m \ge n$ is there since we consider full rank matrices. 

Addition, multiplication, and computing multiplicative or additive inverse are field operations of which various implementation options exist.
It is likely that such operations are implemented with even more Toffoli gates (and work qubits), and then the Toffoli depth of the entire circuit is dominated by the Toffoli depth incurred by the certain arithmetic operation.
For example, assuming the multiplication depth is the dominant factor and the Toffoli depth of a single multiplier is $O(l \lceil\log_{2}p\rceil^2)$, the total Toffoli depth would read $O(l \lceil\log_{2}p\rceil^2 n^2 )$.

In $\mathrm{GF}(2)$, the arithmetic operations are greatly simplified giving the following result:

\begin{corollary}
	For a matrix $\mathbf{A} \in \mathrm{GF}(2)^{m \times n}$, $m \ge n$, Algorithm\;\ref{alg:main-algorithm} is reversibly implementable in which Toffoli gate count and depth are upper bounded by $O(mn^{2})$ and $O(m\log_{2}n + n(\log_{2}m)^{4})$, respectively.
	\label{cor:main_GF2}
\end{corollary}

\begin{figure}[htbp]
	\centering
	\includegraphics[width=0.98\textwidth]{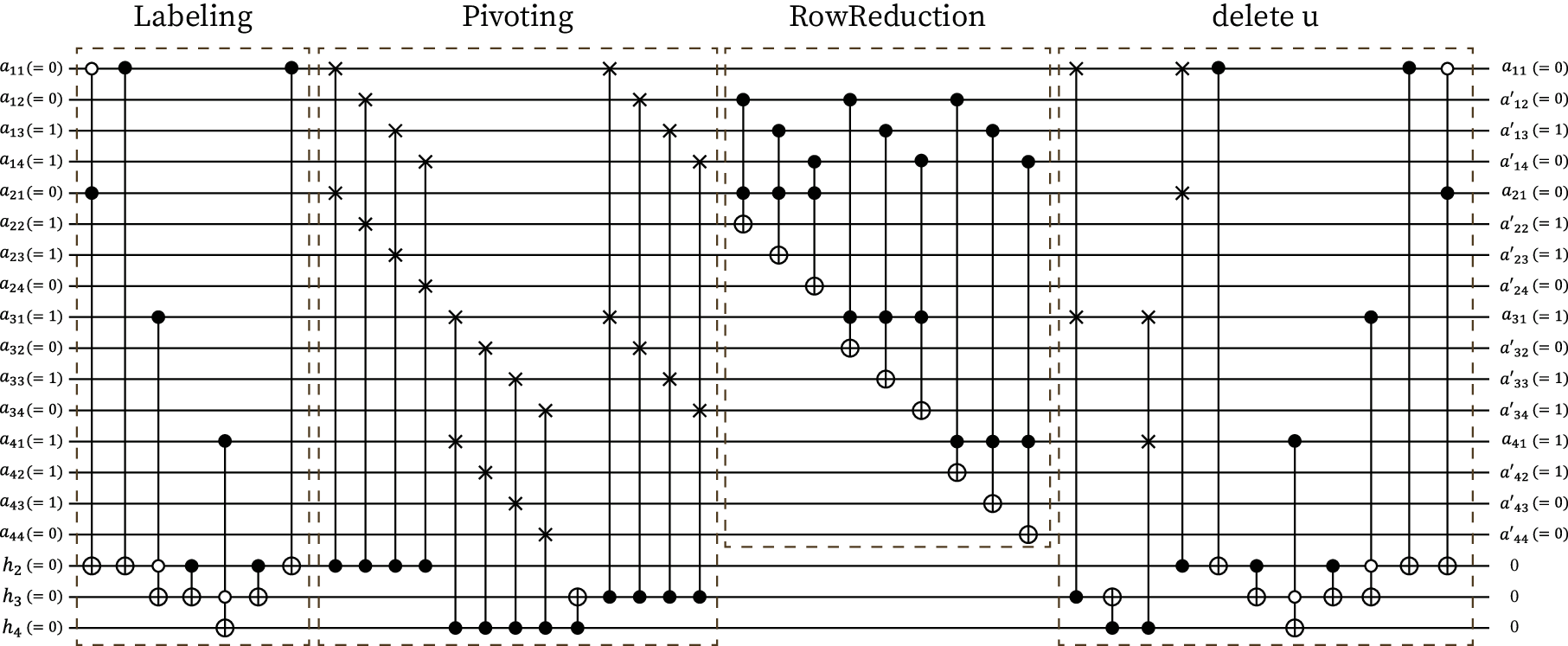}
	\caption{A serial quantum circuit for the first iteration of Algorithm\;\ref{alg:main-algorithm}.}
	\label{fig:GE_example}
\end{figure}

Non-asymptotic bounds are given in Appendix~\ref{app:app-upper-bound}.
As an example, consider the following matrix:
\begin{align}\label{eq:example-4by4}
	\mathbf{A} = \left(
	\begin{matrix}
		a_{11} & a_{12} & a_{13} & a_{14}\\
		a_{21} & a_{22} & a_{23} & a_{24}\\
		a_{31} & a_{32} & a_{33} & a_{34}\\
		a_{41} & a_{42} & a_{43} & a_{44}
	\end{matrix}\right)
	=
	\left(\begin{matrix}
		0 & 0 & 1 & 1\\
		0 & 1 & 1 & 0\\
		1 & 0 & 1 & 0\\
		1 & 1 & 1 & 0
	\end{matrix}\right).
\end{align}
A serial quantum circuit for Algorithm~\ref{alg:main-algorithm} on $\mathbf{A}$ is illustrated in Figure~\ref{fig:GE_example} and the corresponding unitary transformations are given in Eq.\;(\ref{eq:GE_example}).
In Eq.\;(\ref{eq:GE_example}), $\vec{h}$ denotes a vector $(h_{2},h_{3},h_{4})$.
A design for each part of the figure is described in Section\;\ref{sec:4}.

\vspace{-30pt}
\begin{align}\label{eq:GE_example}
		\mathbf{A} &=
		\left(\begin{matrix}
			0 & 0 & 1 & 1\\
			0 & 1 & 1 & 0\\
			1 & 0 & 1 & 0\\
			1 & 1 & 1 & 0
		\end{matrix}\right)
\!\!\!\!\!&&\overset{\rotatebox{90}{\tiny Labeling}}{\mapsto}
		\left(\begin{matrix}
			0 & 0 & 1 & 1\\
			0 & 1 & 1 & 0\\
			1 & 0 & 1 & 0\\
			1 & 1 & 1 & 0
		\end{matrix}\right)
\!\!\!\!\!&&\overset{\rotatebox{90}{\tiny Pivoting}}{\mapsto}
		\left(\begin{matrix}
			\textbf{1} & \textbf{0} & \textbf{1} & \textbf{0}\\
			0 & 1 & 1 & 0\\
			\textbf{0} & \textbf{0} & \textbf{1} & \textbf{1}\\
			1 & 1 & 1 & 0
		\end{matrix}\right)
\!\!\!\!\!&&\overset{\rotatebox{90}{\tiny RowRed.}}{\mapsto}
		\left(\begin{matrix}
			1 & 0 & 1 & 0\\
			0 & 1 & 1 & 0\\
			0 & 0 & 1 & 1\\
			1 & 1 & \textbf{0} & 0
		\end{matrix}\right)
\!\!\!\!\!&&\overset{\rotatebox{90}{\tiny delete u}}{\mapsto}
		\left(\begin{matrix}
			\textbf{0} & 0 & 1 & 0\\
			0 & 1 & 1 & 0\\
			\textbf{1} & 0 & 1 & 1\\
			1 & 1 & 0 & 0
		\end{matrix}\right) \nn
  		\vec{h}
&=~~
		\left(\begin{matrix}
			0 & 0 & 0
		\end{matrix}\right)	
\!\!\!\!\!&&\mapsto~~
		\left(\begin{matrix}
			0 & \textbf{1} & 0
		\end{matrix}\right)	
\!\!\!\!\!&&\mapsto~~
		\left(\begin{matrix}
			0 & 1 & 0
		\end{matrix}\right)	
\!\!\!\!\!&&\mapsto~~
		\left(\begin{matrix}
			0 & 1 & 0
		\end{matrix}\right)	
\!\!\!\!\!&&\mapsto~~
		\left(\begin{matrix}
			0 & \textbf{0} & 0
		\end{matrix}\right)	
\end{align}

\section{Implementations}\label{sec:4}

\subsection{Labeling}\label{sec:labeling}
The purpose of $\mathsf{Labeling}$ is to find the pivot index of $\mathbf{A} \in \text{GF}(p^l)^{m \times n}$.
Be reminded that the pivot index $u$ is an integer, $1\le u \le m$.
At this stage the pivot index is written in unary format, for example for $m=5$, we store $u=1$ as $(0, 0, 0, 0, 1)$ and $u=4$ as $(0, 1, 0, 0, 0)$.

\begin{figure}[htbp]
	\centering
	\includegraphics[width=\textwidth]{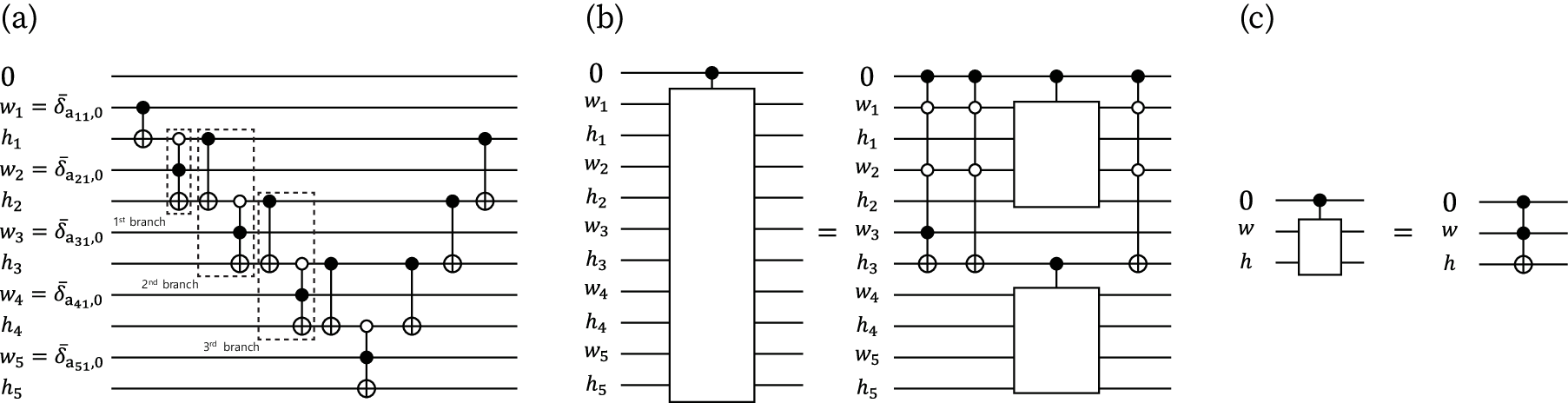}
	\caption{
  (a) A serial implementation of $\mathsf{Labeling}$. Here, $\bar{\delta}_{a,b} = 1$ if $a \ne b$.
  (b) A recursive block diagram for the parallel implementation of $\mathsf{Labeling}$.
  (c) A circuit for the smallest block.
	}
	\label{fig:labeling}
\end{figure}

We need an index register of $m-1$ zeroed qubits, but for now let there be $m$ index qubits and let us denote each qubit by $h_i$, $i\in \{1,...,m\}$.
In addition, one more work register of $m$ zeroed qubits denoted by $w_i$, $i\in \{1,...,m\}$ is also introduced.
For each row $i$, test if $a_{i1}$ is zero or not and then transform $w_i$ as follows:\footnote{This work register is introduced to hold non-zeroness of the first column elements, thus in $\text{GF}(2)$ it is unnecessary.}
\begin{align}
  w_i =
  \left\{
    \begin{array}{ll}
      0, & \hbox{~if $a_{i1}=0$,} \\
      1, & \hbox{~if $a_{i1}\ne 0.$}
    \end{array}
  \right.
  \label{eq:labeling-work}
\end{align}
It can be straightforwardly implemented by using $C^{\alpha} \! X$, $\alpha=l \lceil \log_2 p\rceil$ gates in which the Toffoli count and depth are $O(\alpha (\log_2\alpha)^2 )$ and $O\big( (\log_2 \alpha)^3 \big)$\;\cite{Claudon2024}.
For $m$ rows, the Toffoli count and depth read, $O(m \alpha (\log_2\alpha)^2 )$ and $O\big( (\log_2 \alpha)^3 \big)$, respectively.

For a pivot index $u$, the next goal is to ensure $h_u = 1$ and all the other index qubits stay zero.
Below a few first steps are illustrated, and then a general description will be given.
First, test if $w_1$ is zero or not and then transform $h_1$ as follows:
\begin{align*}
  h_1 =
  \left\{
    \begin{array}{ll}
      0, & \hbox{~if $w_1=0$,} \\
      1, & \hbox{~if $w_1=1,$}
    \end{array}
  \right.
\end{align*}
which can be done by a CNOT gate.
Now we branch (the first branch in Fig.\;\ref{fig:labeling}).
If $w_1 = 1$, then the pivot index $u=(h_m, h_{m-1} ,\ldots, h_2, h_1)=(0,0,\ldots,0,1)$ is to be stored and nothing should be done on the index register afterwards.
If $w_1 = 0$, check if $w_2 = 1$ and if so, let $h_2 = 1$.
It can be done with one Toffoli gate as shown in Fig.\;\ref{fig:labeling}.
The index register should maintain its state afterwards.

Let us investigate one more branch (the second branch in Fig.\;\ref{fig:labeling}).
If $w_1 = w_2 = 0$, we then need to check if $w_3=1$.
Notice at this point that $h_1 \oplus h_2 = 0$ \emph{if and only if} $w_1 = w_2 = 0$.
Thus we may conditionally copy $w_3$ into $h_3$ depending on $h_1 \oplus h_2$.
By observation, it can be implemented by one CNOT and one Toffoli gates as in Fig.\;\ref{fig:labeling}.
At this point, we correctly store $h_3$ but there is possibility that $h_2$ is wrongly changed if $h_1$ was $1$.
This is due to the fact that we store $h_1 \oplus h_2$ in $h_2$ index qubit, meaning that if $h_1=1$, and then $h_2$ also becomes $1$.
To resolve the issue, we need to reverse $h_1 \oplus h_2$ operation once the last index is obtained.

Continuing the construction, one can see that $h_1 \oplus h_2 \oplus \cdots \oplus h_i = 0$ \emph{if and only if} $w_1 = w_2 = \cdots = w_i = 0$.
Let $H_f := \bigoplus_{i=1}^f h_i$.
It is not difficult to see that $H_f$ can be implemented by $f-1$ CNOT gates on $h_1 , \ldots, h_f$ qubits.
Except $h_1$, we have a general expression
\begin{align}\label{eq:label-rule}
  h_i = \overline{H_{i-1}} \wedge w_i \; .
\end{align}
Given $H_{i-1}$ and $w_i$, a quantum circuit implementing $h_{i}$ thus requires one Toffoli gate for AND operation.
Since $H_{i}$ is given by $H_{i-1} \oplus h_{i}$, a CNOT gate follows.
Figure\;\ref{fig:labeling} shows how the construction goes.

The Toffoli depth in serial implementation of Eq.\;(\ref{eq:label-rule}) for $m$ rows is $m-1$.
Now we can always parallelize it to be $O\big( (\log_2 m)^4 \big)$.
Notice that $h_i$ and $H_i$ can also be written as,
\begin{align}\label{eq:label-parallel}
  h_i &= (\overline{w}_1 \wedge \overline{w}_2 \wedge \cdots \wedge \overline{w}_{i-1}) \wedge w_i \,,
\nn
  H_i &= H_{i -1} \oplus h_i = \overline{(\overline{w}_1 \wedge \cdots \wedge \overline{w}_{i-1})} \oplus h_i\, .
\end{align}
Equation\;(\ref{eq:label-parallel}) can be exploited to build a recursive structure\;\cite{Gid25Decompose}.
Consider halving the $m$ rows into two subsets, $R_1 = \{1,\ldots,m/2-1\}$ and $R_2 = \{m/2,\ldots, m \}$.
(Let $m$ be power-of-two for simplicity.)
In $R_1$, we are to obtain $h_1$ through $h_{m/2-1}$ plainly.
In $R_2$, we begin by $h_{m/2} = (\overline{w}_1 \wedge \cdots \wedge \overline{w}_{m/2-1}) \wedge w_{m/2}$ which can be implemented by a $C^{m/2} \! X$ gate.
Similarly, $H_{m/2}$ is implemented by one $C^{(m/2-1)} \! X$ gate and one NOT gate.
From that point on, remaining index digits are plainly computed.
Since a $C^{m/2} \! X$ gate can be implemented by $O\big( (\log_2 \tfrac{m}{2})^3 \big)$\;\cite{Claudon2024}, halving the rows would result in the Toffoli depth of $\tfrac{m}{2}-2 + O\big( (\log_2 \tfrac{m}{2})^3 \big)$.
Similarly, the subcircuit on $R_1$ and $R_2$ can each be halved to reduce the depth to be $\tfrac{m}{4}-2 + O\big((\log_2 \tfrac{m}{2})^3 \big) + O\big( (\log_2 \tfrac{m}{4})^3 \big)$.
Continuing this way, the Toffoli depth of the recursive structure is bounded by $O\big( (\log_2 m)^4 \big)$.
Including implementation of Eq.\;(\ref{eq:labeling-work}), the Toffoli count and depth of $\mathsf{Labeling}$ read $O\big( m\alpha (\log_2 \alpha)^2 + m (\log_2 m)^3 \big)$ and $O\big( (\log_2 \alpha)^3 + (\log_2 m)^4 \big)$, respectively.


There is a simple trick to remove one index qubit, which is briefly covered in Appendix~\ref{sec:app-labeling}.
Intuitively, since one of the unary expression can be replaced by zero-only string, we replace $000\ldots01$ by $000\ldots0$ removing one bit.
Figure\;\ref{fig:labeling} however includes the removable bit ($h_1$) for easier understanding. 
The work register can in principle get cleaned at the end, but we will use it later.



\subsection{Pivoting}\label{sec:pivoting}
The purpose of $\mathsf{Pivoting}$ is to swap the first row and the row labeled by the pivot index.
After $\mathsf{Pivoting}$, the output matrix $\mathbf{A}$ is guaranteed that $a_{11} \ne 0$ (assuming $\mathbf{A}$ is full rank).

\begin{figure}[htbp]
	\centering
	\includegraphics[width=0.25\textwidth]{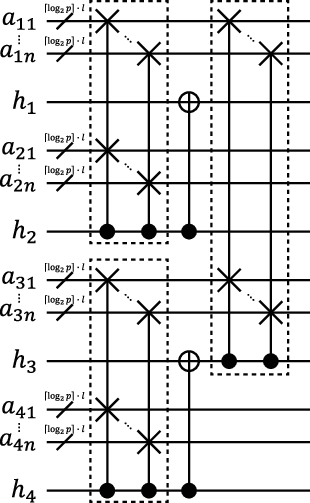}
	\caption{
		A reversible implementation of $\mathsf{Pivoting}$. Dashed boxes mark parallelizable subcircuits by using borrowed qubits. }
	\label{fig:pivoting}
\end{figure}

It has a tournament-like structure.
In the first round, two consecutive rows are paired, for example the first and the second rows are paired, the third and the fourth rows are paired, and so on.
Let $h_i$ be the associated index qubit for the $i$th row (that has been introduced in $\mathsf{Labeling}$).
For the first pair, we swap the two rows if $h_2 = 1$.
It can be carried out by $n\cdot \lceil \log_{2} p \rceil \cdot l$ Fredkin gates as shown in Fig.\;\ref{fig:pivoting}.
After the swap operation, the index information is merged, $h_1 \gets h_1 \oplus h_{2}$.
The other pairs in the first round undergo the same procedure.
Notice that the swap actually takes place in a single pair, since $h_u$ is 1 for the $u$th row only.

In the next round, two every other rows are paired; the first and the third rows, and so on.
It should be apparent how the design goes.
One minor practical point is that in actual implementation, we do not need $h_1$ and thus all the relevant gates, since $h_i$ determines if $i$th row is swapped with \emph{the first row}.
(However, such gates are described in Fig.\;\ref{fig:pivoting} to help understand the construction.)

Once the final round of swapping is completed, the first row is guaranteed that $a_{11} \ne 0$.
The Toffoli depth is estimated below ($\alpha:=l \lceil \log_2 p\rceil$).



\begin{proposition}
	Given qubits encoding a matrix $\mathbf{A} \in \mathrm{GF}(p^{l})^{(m-j') \times (n-j')}$ and its pivot index, and $\alpha (m-j') j'$ borrowed qubits\footnote{It can be thought of as a qubit that is in unknown pure or mixed state.}, $\mathsf{Pivoting}$ is reversibly implementable in which the Toffoli count and depth are bounded by $2(\alpha(n-j') +1) (m-j'-1)$ and  $\frac{2 (n-j')}{j'} + 3\lceil \log_{2}(m-j') \rceil$.
	\label{prop:parallel_pivot}
\end{proposition}

Proof is given in Appendix~\ref{sec:app-pivoting}.

\subsection{Row reduction}
Let $\mathbf{A} \in \text{GF}(p^l)^{m \times n}$ be an input to $\mathsf{RowReduction}$.
A work register of size $m$ is employed of which the qubits are denoted by $\{w_1, \ldots, w_{m}\}$.
Each work qubit $w_i$ is written 1 if and only if $a_{i1}\ne 0$.\footnote{With an appropriate care in $\mathsf{Labeling}$ and $\mathsf{Pivoting}$, the very register introduced in $\mathsf{Labeling}$ can still be used without involving any gate here.}

\begin{align}
	\left(\begin{matrix}
		a_{11} & a_{12} & \cdots & a_{1n}\\
		a_{21} & a_{22} & \cdots & a_{2n}\\
		\vdots & \vdots & \ddots & \vdots\\
		a_{m1} & a_{m2} & \cdots & a_{mn}
	\end{matrix}\right)
	&\underset{\text{Phase1}}{\longmapsto}
	\left(\begin{matrix}
		a_{11} & a_{12} / a_{11} & \cdots & a_{1n} / a_{11}\\
		a_{21} & a_{22} / a_{21} & \cdots & a_{2n} / a_{21}\\
		\vdots & \vdots & \ddots & \vdots\\
		a_{m1} & a_{m2} / a_{m1} & \cdots & a_{mn} / a_{m1}
	\end{matrix}\right)
\nn	&\underset{\text{Phase2}}{\longmapsto}
	\left(\begin{matrix}
		a_{11} & a_{12} / a_{11} & \cdots & a_{1n} / a_{11}\\
		a_{21} & \tfrac{a_{22}}{a_{21}} - \tfrac{a_{12}}{a_{11}} \cdot w_{2} & \cdots & \tfrac{a_{2n}}{a_{21}} - \tfrac{a_{1n}}{a_{11}} \cdot w_{2}\\
		\vdots & \vdots & \ddots & \vdots\\
		a_{m1} & \tfrac{a_{m2}}{a_{m1}} - \tfrac{a_{12}}{a_{11}} \cdot w_{m} & \cdots & \tfrac{a_{mn}}{a_{m1}} - \tfrac{a_{1n}}{a_{11}} \cdot w_{m}
	\end{matrix}\right) \enspace.
\label{eq:row-reduction}
\end{align}


Row reduction takes place in two phases.
In the first phase, each row vector $(a_{i1} ~ a_{i2} ~ \cdots ~ a_{in})$ is turned into $(a_{i1} ~ a_{i2}/a_{i1} ~ \cdots ~ a_{in}/a_{i1})$.
It requires computing the multiplicative inverse of $a_{i1}$ and conditionally (if $w_i = 1$) multiplying $a_{ij}$ by $a_{i1}^{-1}$ for all $j \in \{2,\ldots, n\}$.

In the second phase, the first row is conditionally subtracted from $i$th row for all $i \in \{2, \ldots, m\}$, except the first column elements.
Let the resulting matrix of the first phase be $\mathbf{ A}'$ and let $a'_{ij}$ be its $i$th row, $j$th column element.
For the $i$th row, check if $a'_{i1}$ is zero or not which is easily read off from $w_i$.
If $w_i \ne 0$, and then do $ a'_{ij} \gets a'_{ij} - a'_{1j}$ for all $j \in \{2, \ldots, n\}$.

The above procedures involve arithmetic operations including multiplications, additions, and computing inverses.
Equation\;(\ref{eq:row-reduction}) shows how the two phases work. 
There are various options for implementing arithmetic operations where each option may have distinct time and space complexities.


\begin{proposition}
	Given qubits encoding a matrix $\mathbf{A} \in \mathrm{GF}(p^{l})^{(m-j') \times (n-j')}$ and its pivot index, and $\alpha (m-j') j' + \alpha (n-j'-1) j'$ borrowed qubits, $\mathsf{RowReduction}$ is reversibly implementable in which
	the multiplication count and depth are bounded by $4(m-j')(n-j'-1)$ and $4 \left\lceil \tfrac{n-j'-1}{j'+1} \right\rceil$,
	the multiplicative inverse count and depth are $2(m-j')$ and 2,
	the controlled addition count and depth are bounded by $4(m-j'-1)(n-j'-1)$ and $4 \left\lceil \tfrac{m-j'-1}{j'+1} \right\rceil$, and
	the additive inverse count and depth are bounded by $2(m-j'-1)(n-j'-1)$ and $2 \left\lceil \tfrac{n-j'-1}{j'+1} \right\rceil$
	\label{prop:row-reduction}
\end{proposition}

\begin{remark}
	For $j=1$ in Algorithm~\ref{alg:main-algorithm}, $\mathsf{RowReduction}$ is reversibly implementable in which the number of multiplications, multiplicative inverses, and controlled additions are $2m(n-1)$, $2m$, $(m-1)(n-1)$, and the depth of each is $2(n-1)$, $2$, and $m-1$, respectively.
    Additive inverse is not computed at $j=1$.
	\label{rem:row-reduction}
\end{remark}

Proofs are given in Appendix~\ref{sec:app-rowred}.

\subsection{Deleting the pivot index}
The purpose of deleting the pivot index $u_{j}$ is to limit the use of work (garbage) space by restoring it to zero.
This is essentially reversing $\mathsf{Pivoting}$ only for the first column, followed by fully reversing $\mathsf{Labeling}$.
The asymptotic complexity does not change, and even the exact complexity is not much affected as the heavier operation $\mathsf{Pivoting}$ is only partially reversed. 
\subsection{Complexity}\label{sec:complexity}
The number of Toffoli gates and its depth in Gaussian elimination are estimated below asymptotically, as well as those of arithmetic operations.
Non-asymptotic bounds can also be found in Appendix\;\ref{app:app-upper-bound} with messier expressions.

\paragraph*{Proof of Proposition\;\ref{prop:main_par}.}
(\textit{Labeling})
Let $j' = j-1$.
At $j$th iteration of Algorithm\;\ref{alg:main-algorithm}, the Toffoli count and depth of $\mathsf{Labeling}$ are $O\big( (m-j') \alpha (\log_2 \alpha)^2 + (m-j') (\log_2 (m-j'))^3 \big)$ and $O\big( (\log_2 \alpha)^3 + (\log_2 (m-j'))^4 \big)$ as given in Section\;\ref{sec:labeling}.
It iterates $n$ times, therefore the total Toffoli count due to $\mathsf{Labeling}$ is bounded by,
\begin{align*}
    \sum_{j'=0}^{n-1} &O\big( (m-j') \alpha (\log_2 \alpha)^2 + (m-j') (\log_2 (m-j'))^3 \big)
\nn\le
    \,&O\big( n m \alpha (\log_{2}\alpha)^2 + n m (\log_{2} m)^3 \big) \enspace,
\end{align*}
and the total Toffoli depth due to $\mathsf{Labeling}$ is bounded by,
\begin{align*}
    \sum_{j'=0}^{n} O\big( (\log_2 \alpha)^3 + (\log_2 (m-j'))^4 \big)
\le
    O\big( n (\log_{2}\alpha)^3 + n (\log_{2} m)^4 \big) \enspace.
\end{align*}

(\textit{Pivoting}) At $j$th iteration, there exist at least $\alpha (m-j')j'$ qubits available for borrowed qubits, since for the $m-j'$ rows, the first $j'$ columns do not participate the procedure.
Therefore, the Toffoli count and depth of $\mathsf{Pivoting}$ at $j$th iteration are bounded as in Proposition\;\ref{prop:parallel_pivot}, except $j=1$.
At $j=1$, the count and depth are bounded by $(\alpha n +1) (m-1)$ and $(\alpha n +1) \lceil \log_2 m \rceil$ considering the description of $\mathsf{Pivoting}$ in Section\;\ref{sec:pivoting}.\footnote{The work qubits $w$ that tell nonzeroness of the leading elements are also swapped, giving rise to $+1$ as in $(\alpha n +1)$.}
It iterates $n$ times, therefore the total Toffoli count due to $\mathsf{Pivoting}$ is bounded by,
\begin{align*}
    (\alpha n +1) (m-1) + \sum_{j'=1}^{n-1} 2(\alpha(n-j') +1) (m-j'-1)
\le
    O(\alpha n^2 m) \enspace,
\end{align*}
and the total Toffoli depth due to $\mathsf{Pivoting}$ is bounded by,
\begin{align*}
    (\alpha n +1) \lceil \log_2 m \rceil + \sum_{j=1}^{n} \left[ \frac{2 (n-j')}{j'} + 3\lceil \log_{2}(m-j') \rceil \right]
\le
    O(\alpha n \log_2 m) \enspace.
\end{align*}

(\textit{Deleting the pivot index}) Deleting the pivot index involves partially reversing $\mathsf{Pivoting}$ (only the first column) and fully reversing $\mathsf{Labeling}$.
The leading terms of Toffoli count come from reversing $\mathsf{Labeling}$; $O\big( n m \alpha (\log_{2}\alpha)^2 + n m (\log_{2} m)^3 \big)$.
For the Toffoli depth, in addition to the reversing $\mathsf{Labeling}$, partially reversing $\mathsf{Pivoting}$ contributes to the increase in the depth by $O(\alpha \log_2 m )$ (only the first column), therefore the depth in deleting the pivot index is bounded by $O\big( n (\log_{2}\alpha)^3 + n (\log_{2} m)^4 + \alpha \log_2 m \big)$.

(\textit{Row reduction}) At $j$th iteration, there exist at least $\alpha (m-j') j' + \alpha (n-j'-1) j'$ qubits available for borrowed qubits, since for the $m-j'$ rows ($n-j'$ columns), the first $j'$ columns (rows) do not participate the procedure.
Therefore, the cost of $\mathsf{RowReduction}$ at $j$th iteration is bounded as in Proposition\;\ref{prop:row-reduction}, except $j=1$.
At $j=1$, the cost is bounded by Remark\;\ref{rem:row-reduction}.
It iterates $n$ times, therefore the multiplication count and depth due to $\mathsf{RowReduction}$ are bounded by,
\begin{align*}
    &2m(n-1) + \sum_{j'=1}^{n-1} 4(m-j')(n-1-j')
\le
    O(m n^2) ,
\nn
    &2(n-1) + \sum_{j'=1}^{n-1} 4 \left\lceil \tfrac{n-1-j'}{j'+1} \right\rceil
\le
    O(n \log_2 n) \enspace,
\end{align*}
where the first term in both lines comes from Remark~\ref{rem:row-reduction} and the remaining terms are from iterating $j'$ for Proposition~\ref{prop:row-reduction}.
Similarly, the multiplicative inverse count and depth are bounded by
\begin{align*}
    2m + \sum_{j'=1}^{n-1} 2(m-j')
\le
    O(m n) ,
\qquad
    2 + \sum_{j'=1}^{n-1} 2
\le
    O(n) \enspace,
\end{align*}
and the controlled addition count and depth are bounded by
\begin{align*}
	&(m-1)(n-1) + \sum_{j'=1}^{n-1} 4(m-1-j')(n-1-j')
\le
	O(m n^2) ,
\nn
	&m-1 + \sum_{j'=1}^{n-1} 4 \left\lceil \tfrac{m-1-j'}{j'+1} \right\rceil 
\le
	O(m \log_2 n) \enspace,
\end{align*}
and the additive inverse count and depth are bounded by
\begin{align*}
	\sum_{j'=1}^{n-1} 2(m-1-j')(n-1-j')
\le
	O(m n^2) ,
\qquad
	\sum_{j'=1}^{n-1} 2 \left\lceil \tfrac{n-1-j'}{j'+1} \right\rceil
\le
	O(n \log_2 n) \enspace.
\end{align*}
$\qed$

\section{Discussions}\label{sec:5}
Compared with previous works on quantum implementation of Gaussian elimination in $\text{GF}(2)$\;\cite{GE21,GE23}, there are two main points we would like to stress on.

The first is that the design developed in this work is a generalization of the previous ones as it works in any finite fields, not just in $\text{GF}(2)$.
Working in $\text{GF}(2)$ has many advantages in various aspects, for example addition is a simple XOR, or its non-zeroness is trivially read off.

The second is that the two previous designs have their own strong and weak points in time and space aspects as shown in Table\;\ref{tab:complexity-comparison}.
We managed to optimize their respective weak points resulting in matching classical complexity while limiting the use of extra space.
The detailed improvement points are not clearly apparent as component-by-component comparisons are not immediately describable.
In the main text, rather than trying to compare with previous works, we have focused on explaining the design criteria in each part.

One minor note is that there still exist windows for further improvements upon the circuit we developed, for example the multiplication depth could be shortened since the same multiplicative inverse element is multiplied to $n-1$ different elements in the row reduction.
Such optimizations might be worth investigating if Gaussian elimination is to be implemented on a realistic hardware.



\bibliographystyle{splncs04}
\bibliography{reference}


\newpage
\appendix
\newpage
\section{Parallelization with borrowed qubits}\label{sec:app-ParBow}
We describe a parallelization method based on \emph{borrowed qubits}, motivated by Gidney’s approach to circuit decomposition~\cite{Gid25Decompose}.
The purpose of this method is to execute multiple overlapping operations in parallel while restoring every borrowed qubit to its original value at the end of the computation.
In our setting, the relevant target operations are the controlled swap operations used in $\mathsf{Pivoting}$ and the arithmetic operations used in \textsf{RowReduction}.

The basic idea is common to all cases.
When several operations share the same control or one of the same input values, we temporarily transfer the relevant information to borrowed qubits, use those qubits in parallel operations, and then apply correction and restoration steps so that the overall transformation is identical to the intended computation.
As a result, the borrowed qubits serve only as temporary workspace and are restored to their initial states after the parallelized procedure.

\subsection{Transferring of the shared value to borrowed qubits}
\label{app:distributing}
\begin{figure}[H]
	\centering
	\includegraphics[width=0.90\textwidth]{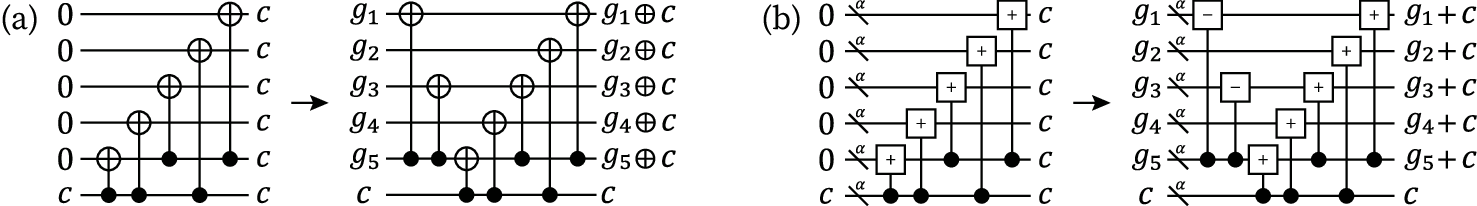}
	\caption{
		Transfer of a value $c$ to auxiliary qubits, either clean or borrowed.
		(a) illustrates the CNOT-based transfer of $c$: to clean qubits in the left circuit, and to borrowed qubits in the right circuit, whose initial values are $g_{1}, \ldots, g_{5}$.
		(b) illustrates the analogous transfer based on addition: to clean qubits in the left circuit, and to borrowed qubits in the right circuit, again with initial values $g_{1}, \ldots, g_{5}$.
		The $(+)$ and $(-)$ boxes indicate that the controlled value is added to or subtracted from the target value, respectively.
	}
	\label{fig:Expanding}
\end{figure}

To parallelize operations using borrowed qubits, the repeatedly referenced qubits (data qubits hereafter) must first be transferred to the borrowed qubits.
When borrowed qubits are used for parallelization, it is important to transfer the data qubit in logarithmic depth so that the dominant cost remains in the target operations.
To achieve logarithmic depth, this transfer is performed recursively using previously spread information, together with correction steps that ensure the intended overall transformation.
By observation, for example in Fig.~\ref{fig:Expanding}, transferring a data value to $g$ borrowed qubits requires $2g - \left\lceil \log_2 (g+1) \right\rceil$ operations and has depth bounded by $2\left\lceil \log_2 (g+1) \right\rceil$.
Briefly, the transferring requires at least $g$ operations.
Moreover, the correction step requires $g-\left\lceil \log_{2}(g+1)\right\rceil$ operations, since the $\left\lceil \log_{2}(g+1)\right\rceil$ transfers originating directly from the original value $c$ do not need to be corrected.
Therefore, the total number of operations is $2g-\left\lceil \log_{2}(g+1)\right\rceil$.
Similarly, the transferring has depth $\left\lceil \log_{2}(g+1)\right\rceil$, and the correction step contributes an additional depth of $\left\lceil \log_{2}g\right\rceil-1$ when $g \neq 1$.
Hence, the overall depth is $1$ for $g=1$ and $\left\lceil \log_{2}(g+1) \right\rceil + \left\lceil\log_{2}g\right\rceil - 1 $ for $g\neq1$ and both are bounded by $2\left\lceil \log_{2}(g+1)\right\rceil$.
The following figures illustrate how this logarithmic distribution procedure is implemented.
\vspace{-10pt}

\subsection{Parallelization of controlled swap operations for $\mathsf{Pivoting}$}
\label{app:parallel-pivoting}

\begin{figure}[H]
	\centering
	\includegraphics[width=0.55\textwidth]{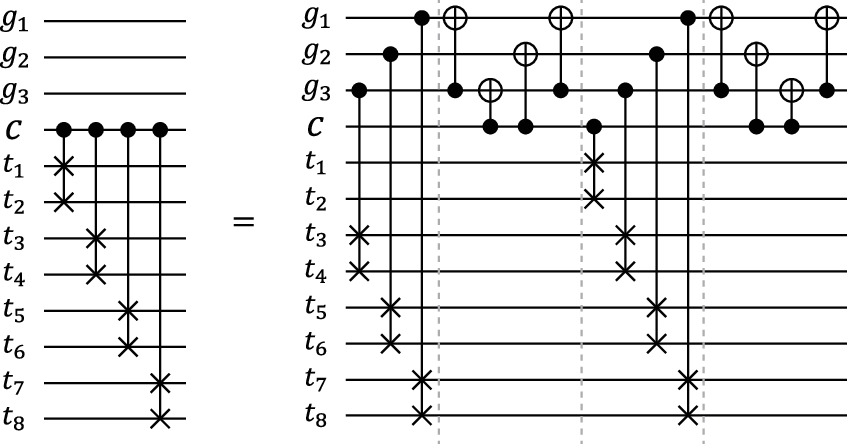}
	\caption{
		Example illustrating the parallelization of four controlled swap operations sharing the same control qubit.
		The parallelized circuit has controlled swap depth 2.
	}
	\label{fig:Parallelization_Swap_witn_BorrowedQubit}
\end{figure}

We first consider controlled swap operations conditioned on a single control qubit.
Such operations can be parallelized by transferring the control value onto a borrowed qubit and then using that borrowed qubit for other controlled swaps.
Because the borrowed qubit in general initially contain an unknown value, an additional correction step is necessary to ensure that the overall transformation coincides exactly with the desired controlled swap.
This correction can be implemented by inserting an additional controlled swap operation before the main operation.
Accordingly, parallelizing a single controlled swap operation requires:
\begin{itemize}
	\item two controlled swap operations, and
	\item two CNOT gates for transferring and uncomputing the control value.
\end{itemize}
An example is shown in Fig.~\ref{fig:Parallelization_Swap_witn_BorrowedQubit}.
By observation, if the original controlled swap depth is $d$, then using $g$ borrowed qubits reduces the depth bounded by $2\left\lceil \frac{d}{g+1} \right\rceil$.
By Appendix~\ref{app:distributing}, the corresponding CNOT count and depth are bounded by $4g$ and $4\left\lceil \log_2 (g+1) \right\rceil$, respectively.

\subsection{Parallelization of Arithmetic Operations for \textsc{RowReduction}}
\label{app:parallel-rowreduction}
The \textsf{RowReduction} procedure consists of two phases.
In the first phase, quantum--quantum multiplications sharing the same input value appear, whereas in the second phase, controlled quantum--quantum additions either sharing the same input or control value appear.
We note that the target arithmetic operations can be written in the form $t + v\tau$.

\begin{figure}[htbp]
	\centering
	\includegraphics[width=0.6\textwidth]{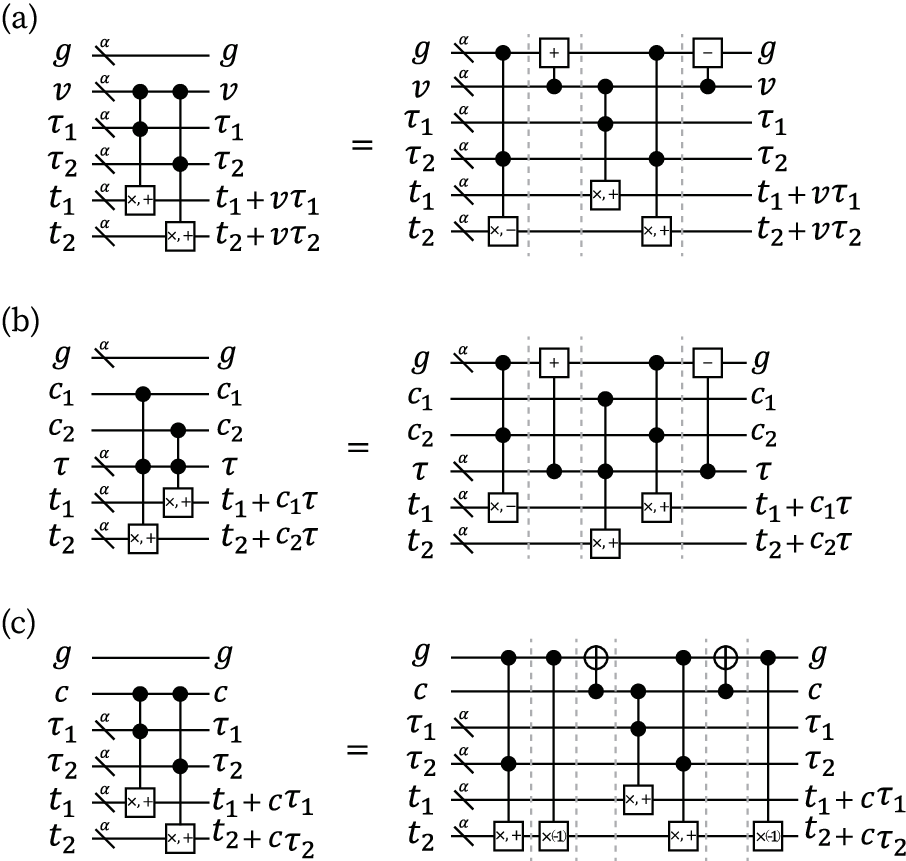}
	\caption{
		(a), (b) and (c) illustrate the parallel implementations of the arithmetic operations in the first and second phase of \textsf{RowReduction} using borrowed qubits.
		(a) shows the parallelization of the quantum--quantum multiplication in the first phase, while (b) and (c) show the parallelization of the controlled quantum--quantum addition in the second phase.
		(b) corresponds to the case of overlapping columns, and (c) to the case of overlapping rows.
		The $(\times,+)$ box denotes that the product of the two multiplicands is added to the target value, whereas the $(\times,-)$ box denotes that the corresponding product is subtracted from the target value.
		Moreover, the $\big(\times(-1)\big)$ box denotes a controlled multiplication by $-1$ on the target value, thereby mapping it to its additive inverse.
	}
	\label{fig:Parallelization_witn_BorrowedQubit}
\end{figure}

The parallelization of the operation $t+v\tau$ is divided into three cases.
The operations reference the same
\begin{itemize}
	\item[(1)] $v$ where $v,\tau \in \mathrm{GF}(p^{l})$, 
	\item[(2)] $v$ where $v \in \{0,1\}, \tau \in \mathrm{GF}(p^{l})$,
	\item[(3)] $\tau$ where $v \in \{0,1\}, \tau \in \mathrm{GF}(p^{l})$.
\end{itemize}
Figure~\ref{fig:Parallelization_witn_BorrowedQubit} summarizes the resulting constructions for parallelization with borrowed qubits.

The basic strategy is based on the transformation
\begin{align*}
	t \;\mapsto\; t - g\tau + (g+v)\tau \;=\; t + v\tau .
\end{align*}
However, in Case~(2) where $v \in \{0,1\}$ is the shared value, one must compute $g+v$ in $\mathbb{Z}$.
Since $v$ is a single qubit in this setting, it is preferable to realize this step by a CNOT gate rather than by a full addition operation.
Further details are provided in Eq.~\eqref{eq:parAddition_with_additiveInv}, and the relevant paragraphs.

\subsubsection{First Phase: Parallelization of Quantum--Quantum Multiplications}
We consider out-of-place quantum--quantum multiplications that reference the same multiplicand value.
These operations can be parallelized by transferring the multiplicand value into borrowed qubits and then using the borrowed qubits for other multiplications.
Because the borrowed qubits may initially contain an unknown value, a correction step is necessary.
The correction is implemented by inserting an additional quantum--quantum multiplication before the main multiplication.
Accordingly, parallelizing a single quantum--quantum multiplication requires:
\begin{itemize}
	\item two quantum--quantum multiplications, and
	\item two quantum--quantum additions for transferring and uncomputing.
\end{itemize}
An illustrative circuit is shown in Fig.~\ref{fig:Parallelization_witn_BorrowedQubit}(a).
By observation, if the original multiplication depth is $d$, then using $\alpha g$ borrowed qubits reduces the depth to $2\left\lceil \frac{d}{g+1} \right\rceil$.
By Appendix~\ref{app:distributing}, the corresponding addition/subtraction count and depth are bounded by $4g$ and $4\left\lceil \log_2 (g+1) \right\rceil$, respectively.

\subsubsection{Second Phase: Parallelization of Controlled Quantum--Quantum Additions}
In the second phase, for a matrix $\mathbf{A} \in \mathrm{GF}(p^{l})^{m \times n}$, the main task is to perform subtracting the first row from other rows.
Therefore, elimination with respect to the first row is performed on the remaining $m-1$ rows over the $n-1$ columns.
Accordingly in the serial implementation, it requires $(m-1)(n-1)$ controlled additions.

Main parallelization points are two-folds.
Notice that when a row $\v a = (a_1, ..., a_n)$ is subtracted from another row $\v a' = (a'_1, ..., a'_n)$, $a'_1$ is referenced $n-1$ times.
In the proposed construction, the non-zeroness of the leading element is written on a single work qubit which is referenced $n-1$ times accounting for one of the two main bottlenecks in the second phase.
By sharing the workload of the work qubit, the depth can be reduced.
We will call it parallelization in the row direction.
In addition, also notice that in the proposed construction, the first row is subtracted from all the other rows; $m-1$ times.
By parallelizing the first row, the depth can be reduced, which will be called the parallelization in the column direction.
We discuss these two cases in turn.

\paragraph{Case 1: column direction}
In column direction, we need to consider controlled quantum--quantum additions that reference the same addend value.
In this case, the addend value is transferred to borrowed qubits and then used as an input for one of the parallel controlled addition.
Since the borrowed qubits may initially store an unknown value, an additional correction step is required, which is implemented by inserting one extra quantum--quantum addition before the main controlled addition.
Thus, parallelizing a controlled quantum--quantum addition requires:
\begin{itemize}
	\item two controlled quantum--quantum addition, and
	\item two quantum--quantum additions for transferring and uncomputing.
\end{itemize}
An illustrative circuit is shown in Fig.~\ref{fig:Parallelization_witn_BorrowedQubit}(b).
By observation, if the original controlled addition depth is $d$, then using $\alpha g$ borrowed qubits reduces the depth to $2\left\lceil \frac{d}{g+1} \right\rceil$.
By Appendix~\ref{app:distributing}, the corresponding addition/subtraction count and depth are bounded by $4g$ and $4\left\lceil \log_2 (g+1) \right\rceil$, respectively.

\paragraph{Case 2: row direction}
In row direction, we need to consider controlled quantum--quantum additions that are conditioned on the same control qubit.
In this case, the control value is transferred to borrowed qubits and then used to drive parallel executions.
Similar to column direction, the borrowed qubits may initially store an unknown value, therefore it needs some correction step.
However, in this case, the correction step becomes slightly more involved, as it must be implemented using a CNOT gate rather than an addition.
To recover the exact intended transformation, we use two controlled additive inversions in addition to the copying and uncomputation steps.
The process is described as
\begin{align}
	\ket{g}_{1}\ket{c}_{1}\ket{a}\ket{b}
	\mapsto&~ \ket{g}_{1}\ket{c}_{1}\ket{a}\ket{b+ga}\nn
	\mapsto&~ \ket{g}_{1}\ket{c}_{1}\ket{a}\ket{(-1)^{g}(b+ga)}\nn
	\mapsto&~ \ket{g \oplus c}_{1}\ket{c}_{1}\ket{a}\ket{(-1)^{g}(b+ga)}\nn
	\mapsto&~ \ket{g \oplus c}_{1}\ket{c}_{1}\ket{a}\ket{(-1)^{g}(b+ga) + (g\oplus c)a}\nn
	\mapsto&~ \ket{g}_{1}\ket{c}_{1}\ket{a}\ket{(-1)^{g}(b+ga) + (g\oplus c)a}\nn
	\mapsto&~ \ket{g}_{1}\ket{c}_{1}\ket{a}\ket{(-1)^{g}\big((-1)^{g}(b+ga) + (g\oplus c)a\big)}\nn
	= &~ \ket{g}_{1}\ket{c}_{1}\ket{a}\ket{b + \bigl(g + (-1)^g(g \oplus c)\bigr)a} \nn
	= &~ \ket{g}_{1}\ket{c}_{1}\ket{a}\ket{b + ca},
	\label{eq:parAddition_with_additiveInv}
\end{align}
where the inversions are involved in the second and the sixth maps.

\begin{figure}[H]
	\centering
	\includegraphics[width=0.6\textwidth]{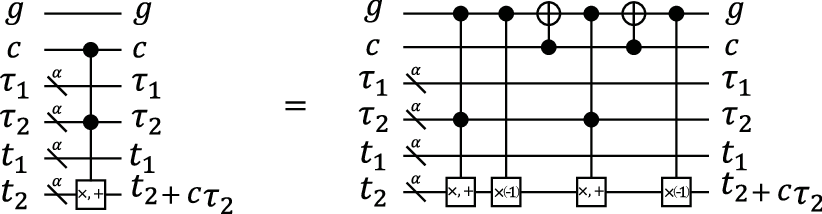}
	\caption{
		Illustration of each map in Eq.~\eqref{eq:parAddition_with_additiveInv}.
		The box notation follows Fig.~\ref{fig:Parallelization_witn_BorrowedQubit}.
	}
	\label{fig:Parallelization_witn_BorrowedQubit2}
\end{figure}

Parallelizing a single controlled quantum--quantum addition in row direction requires:
\begin{itemize}
	\item two controlled quantum--quantum additions,
	\item two controlled additive inversions, and
	\item two CNOT gates for copying and uncomputing the control value.
\end{itemize}
An illustrative circuit is shown in Fig.~\ref{fig:Parallelization_witn_BorrowedQubit}(c).
By observation, if the original controlled addition depth is $d$, then using $g$ borrowed qubits reduces the depth to $2\left\lceil \frac{d}{g+1} \right\rceil$.
Again by Appendix~\ref{app:distributing}, the CNOT count and depth are bounded by $4g$ and $4\left\lceil \log_2 (g+1) \right\rceil$, respectively, while the controlled additive inversion count and depth are $2g$ and $2\left\lceil \frac{d}{g+1} \right\rceil$, respectively.

\section{Details on subroutines}\label{sec:app-ImplDetail}
We propose the details of each algorithm and its complexities.
Moreover, we introduce parallelization version of each procedure and its complexities.

In Algorithm~\ref{alg:main-algorithm}, we omit $\mathbf{w}$ which is defined in Eq.~\eqref{eq:labeling-work} for notational simplicity.
However, we explicitly include $\mathbf{w}$ in each algorithm when describing the procedure.
\subsection{Labeling}\label{sec:app-labeling}
\subsubsection{Algorithm}
The purpose of the $\mathsf{Labeling}$ operation is to encode in unary form the index of the pivot row that will be moved to the top row.
The following is the pseudo algorithm for $\mathsf{Labeling}$.
In the following algorithm, $\delta_{(i,j)}$ is the Kronecker delta defined to be 1 when $i=j$, and 0 when $i \neq j$.
We also define $\bar{\delta}_{(i,j)} := 1 - \delta_{(i,j)}$, equivalently $\bar{\delta}_{(i,j)} = 1$ if $i \neq j$ and $\bar{\delta}_{(i,j)} = 0$ if $i = j$.
Note that $\mathsf{Labeling}$ may return -1 when the given matrix has no non-zero pivot row.
\begin{algorithm}[H]
	\small
	\caption{$\mathsf{Labeling}$}
	\begin{algorithmic}[1]
		\Require $\mathbf{A} \in \text{GF}(p^{l})^{m \times n}$
		\Ensure $u, \mathbf{w} \in \{0,1\}^{m}$
		\Comment{pivot index in unary format}
		\algrule
		\State $u \gets -1$
		\State $\mathbf{w} = (w_{1}, \ldots, w_{m}) \gets (\bar{\delta}_{(a_{11},0)}, \ldots, \bar{\delta}_{(a_{m1},0)})$
		\Comment{Eq.~\eqref{eq:labeling-work}}
		\For{$j \in \{1, \ldots, m\}$}
		\Comment{$j$: row iterator}
			\If{$w_{j} = 1$}
				\Comment{$w_{j} = 1 \iff a_{j1} \neq 0$}
				\State $u \gets j - 1$
				\State \textbf{break}
			\EndIf
		\EndFor
		\State \textbf{return} $u$, $\mathbf{w}$
	\end{algorithmic}
	\label{alg:labeling}
\end{algorithm}

\begin{figure}
	\centering
	\includegraphics[width=0.9\textwidth]{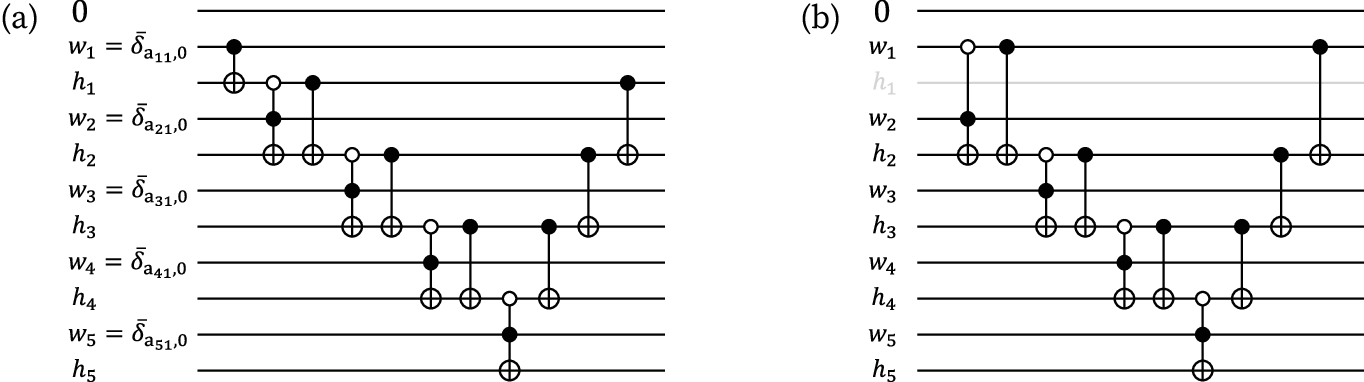}
	\caption{Circuits for $\mathsf{Labeling}$: (a) basic implementation as shown in Fig.~\ref{fig:labeling}, (b) other implementation that saves one qubit.}
	\label{fig:Lab remove one qubit}
\end{figure}

$\mathsf{Labeling}$ operation generates control qubits that determine whether all rows should be moved during the $\mathsf{Pivoting}$ operation.
A straightforward approach is to encode the index of the pivot row in unary format, from which the control qubits can be directly derived. This enables the circuit to function as illustrated in Fig~\ref{fig:labeling}.
Meanwhile, a control qubit for the first row is not required, since no relocation is needed when it serves as the pivot row.
Therefore, the first qubit in the unary format is not used, while the value of unary must be zero when the first row is the pivot.
In addition to the definition of the pivot index presented in the main body, the output of the Labeling procedure can be zero.
In such cases, it indicates that the first row is selected as the pivot.
Moreover, this process is simply achieved by replacing $h_{1}$ with $w_{1}$ in $\mathsf{Labeling}$ operation, and as a result, the circuit operates with $m-1$ qubits instead of $m$.
The resulting circuit can be found in Fig~\ref{fig:Lab remove one qubit}.


\subsubsection{Parallelization}
Motivated by Gidney’s approach to circuit decomposition~\cite{Gid25Decompose}, labeling circuit admits a parallel decomposition as well.
The main idea for parallelizing labeling circuit is halving the circuit with two smaller labeling circuits.
Consider two halved subsets $R_{1} = \{1, \ldots, \lceil m/2\rceil-1 \}$ and $R_{2} = \{\lceil m/2\rceil+1, \ldots, m\}$.
In $R_{1}$, $h_{1}, \ldots, h_{\lceil m/2\rceil -1}$ can be computed plainly.
In $R_{2}$, we begin by $h_{\lceil m/2 \rceil} = (\overline{w}_{1} \land \cdots \land \overline{w}_{\lceil m/2 \rceil - 1}) \land w_{\lceil m/2 \rceil}$ based on Eq.~\eqref{eq:label-parallel} which can be implemented by a $C^{\lceil m/2\rceil}\!X$ gate.
After that, it is changed to $\overline{H}_{\lceil m/2 \rceil}$ by $\overline{H}_{\lceil m/2 \rceil} = h_{\lceil m/2 \rceil} \oplus (\overline{w}_{1} \land \cdots \land \overline{w}_{\lceil m/2\rceil-1})$ which can be computed by a  $C^{\lceil m/2 \rceil - 1}\!X$ gate.
Before computing $R_{2} \backslash \{\lceil m/2 \rceil\} = \{h_{\lceil m/2 \rceil+1}, \ldots, h_{m}\}$, if one of $w_{1}, \ldots, w_{\lceil m/2 \rceil}$ is non-zero, then all values of $h_{\lceil m/2 \rceil+1}, \ldots, h_{m}$ should be zero.
In contrast, if $w_{1}, \ldots, w_{\lceil m/2 \rceil}$ are zero, then $h_{\lceil m/2 \rceil+1}, \ldots, h_{m}$ are computed plainly based on the $w_{\lceil m/2 \rceil+1}, \ldots, w_{m}$ values.
Moreover, we note that $\overline{H}_{\lceil m/2 \rceil} = \overline{w}_{1} \land \cdots \land \overline{w}_{\lceil m/2 \rceil}$ is zero when one of $w_{1}, \ldots, w_{\lceil m/2 \rceil}$ is non-zero.
Therefore, $h_{\lceil m/2 \rceil+1}, \ldots h_{m}$ can be computed by labeling with $w_{\lceil m/2 \rceil+1}, \ldots, w_{m}$ controlled by $\overline{H}_{\lceil m/2 \rceil}$.
After that, $h_{\lceil m/2 \rceil}$ is restored by a $C^{\lceil m/2 \rceil - 1}\!X$ gate.
These parallelization for labeling is applied to the smallest circuit and the examples are shown in Fig~\ref{fig:labeling}.
A detailed algorithm is presented below.
In the following parallelized algorithm, the input is the first column $\mathbf{a} = (a_{1}, a_{2}, \ldots, a_{m}) \in \mathrm{GF}(p^{l})^{m}$ of the matrix $\mathbf{A} \in \mathrm{GF}(p^{l})^{m \times n}$, rather than $\mathbf{A}$ itself.

\begin{algorithm}[H]
	\small
	\caption{$\mathsf{ParallelLabeling}$}
	\begin{algorithmic}[1]
		\Require $c \in \{0, 1\}$, $m' \in \mathbb{Z}$, $\mathbf{a} = (a_{1}, \ldots, a_{m}) \in \text{GF}(p^{l})^{m}$
		\Ensure $u$
		\Comment{pivot index in unary format}
		\algrule
		\State \textbf{if} $c = 1$ \textbf{then return} $0$
		\If {$m=1$}
			\State \Return $m' + \bar{\delta}_{(a_{1},0)} - 1$
		\Else
			\State $c' \gets 0$
			\State \textbf{if} $\big(\bar{\delta}_{(a_{1},0)} \neq 0 \text{ or } \ldots \text{ or } \bar{\delta}_{(a_{\lceil m/2 \rceil - 1},0)} \neq 0\big)$ \textbf{then} $c' \gets 1$
			\State $\mathbf{a}_{1} \gets (a_{1}, \ldots, a_{\lceil m/2 \rceil - 1})$
			\State $\mathbf{a}_{2} \gets (a_{\lceil m/2 \rceil+1}, \ldots, a_{m})$
			\State \Return $\mathsf{ParallelLabeling}(0, \lceil m/2 \rceil-1, \mathbf{a}_{1}) + \mathsf{ParallelLabeling}(c', \lfloor m/2 \rfloor, \mathbf{a}_{2})$
		\EndIf
	\end{algorithmic}
	\label{alg:labeling_paralleized}
\end{algorithm}

\subsection{Pivoting}\label{sec:app-pivoting}
\subsubsection{Algorithm}
\begin{figure}[htbp]
	\centering
	\includegraphics[width=0.7\textwidth]{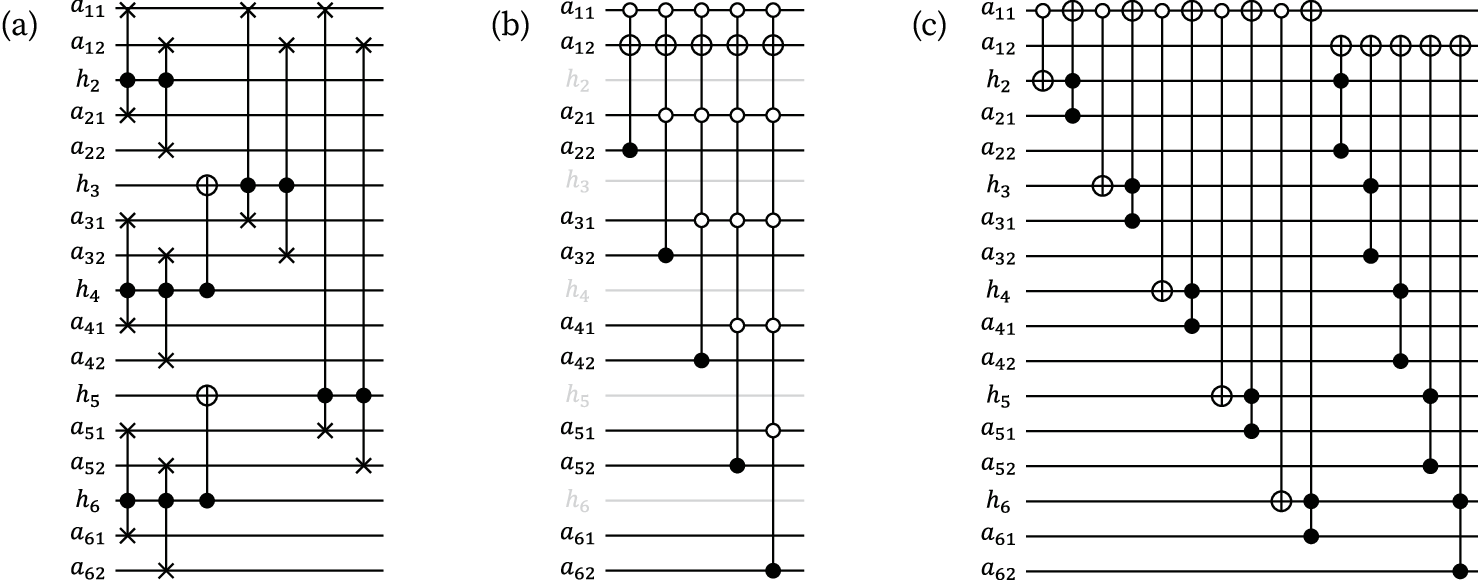}
	\caption{
		(a) is resulting circuit of $\mathsf{Pivoting}$ operation for $\mathbf{A} \in \mathrm{GF}(2)^{6 \times 2}$.
		(b) and (c) are circuits of the counterpart operations in \cite{GE21} and \cite{GE23}, respectively.
		The corresponding circuit in \cite{GE21} doesn't utilize any ancilla qubits but requires multiple controlled gates, while that of \cite{GE23} uses ancilla qubits, which helps to reduce depth of Toffoli gate.
		}
	\label{fig:pivoting_with_others}
	\vspace{-15pt}
\end{figure}

The purpose of the $\mathsf{Pivoting}$ operation is to move the pivot row to the top, based on the value of $u$.
The following is the pseudo algorithm for $\mathsf{Pivoting}$ that works serially.
Since there doesn't exist $\mathbf{w}$ when matrix is defined over $\mathrm{GF}(2)$, pseudo algorithm for $\mathsf{Pivoting}$ is presented with $\mathbf{w}$.
For $\mathrm{GF}(2)$, $\mathsf{Pivoting}$ runs without $\mathbf{w}$ simply.
%
%
\begin{algorithm}[H]
	\small
	\caption{$\mathsf{SerialPivoting}$}
	\begin{algorithmic}[1]
		\Require $\mathbf{A} \in \text{GF}(p^{l})^{m \times n}$, $u$, $\mathbf{w} \in \{0,1\}^{m}$
		\Ensure $\mathbf{A}'' \in \text{GF}(p^{l})^{m \times n}$, $\mathbf{w}'' \in \{0,1\}^{m}$
		\algrule
		\State $\mathbf{A}' \gets (\mathbf{A} \;|\; \mathbf{w})$
		\For{$d \in \{1, \ldots, \lceil \log_{2}m \rceil \}$}
			\State $\Delta \gets 2^{d-1}$
			\For{$j \in \{1, \ldots, \lceil m/2^{d} \rceil\}$}
				\State $r_{2} \gets (m + 1 - \Delta) - (j-1)2^{d}$
				\State $r_{1} \gets \max(r_{2} - \Delta, 1)$
				\If{$u=r_{2}$ and $r_{1} \neq r_{2}$ and $r_{1}, r_{2} \ge 1$}
					\State $\mathbf{A}' \gets \mathsf{swap}(\mathbf{A}', r_{1}, r_{2})$
					\State $u \gets u - \Delta$
				\EndIf
			\EndFor
		\EndFor
		\State $(\mathbf{A}'' \;|\; \mathbf{w}'') \gets \mathbf{A}'$
		\State \textbf{return} $\mathbf{A}'', \mathbf{w}''$
	\end{algorithmic}
	\label{alg:pivoting}
\end{algorithm}
%
$\mathsf{SerialPivoting}$ operation swaps two rows based on the value of label qubits with $m-1$ times.
In the procedure, there exist swap operations that can be parallely executable.
Therefore, $\mathsf{SerialPivoting}$ operation swaps parallelizable rows as much as possible in the same time, which realized with a tournament-like structure.
Furthermore, since the label stores a 1 at the position corresponding to the row to be moved, the associated data must be transferred accordingly in the following parallel swap step (In Algorithm~\ref{alg:pivoting}, value of $u$ is updated.).
This process is implemented using CNOT gates. Conceptually, it can be understood as attaching the label information to the winner in a tournament-style selection, allowing the label to propagate along with the selected value. The resulting circuit is shown in Fig~\ref{fig:pivoting_with_others}.

We now compare different pivoting strategies in Gaussian elimination over $\mathrm{GF}(2)$~\cite{GE21,GE23}.
In our proposed circuit, a row of $n$ elements is swapped a total of $m-1$ times, resulting in $n(m-1)$ controlled swaps.
These swaps can be performed in parallel, leading to a circuit depth of $n \lceil \log_{2}m\rceil$.
In the approach of \cite{GE21}, it performs row additions under the condition that none of the preceding rows serve as the pivot row.
This method requires no additional qubits but involves multiple controlled not gates.
By applying the decomposition method introduced in \cite{Claudon2024}, which utilizes one borrowed qubit\footnote{In the approach of \cite{GE21}, it is available to access to a borrowed qubit.}, the approach in \cite{GE21} results in $O(nm^{2}(\log_{2}m)^{2})$ Toffoli gates and a depth of $O(nm(\log_{2}m)^{3})$.
In the approach of \cite{GE23}, it reduces the complexity significantly by introducing additional qubits.
As in \cite{GE21}, the circuit repeatedly adds rows until the first non-zero row is found, but it stores the corresponding control information using ancilla qubits and restricts the addition operations to the first column.
This allows the remaining columns to be updated in parallel.
Consequently, although the total number of Toffoli gates remains $n(m-1)$, the depth is reduced to $m-1 + \max(n-1,m-1)$.

\vspace{-7pt}
\subsubsection{Parallelization}
Based on the idea for parallelizaion controlled swap with borrowed qubits as in Appendix~\ref{sec:app-ParBow}, $\mathsf{Pivoting}$ can be parallelized in each controlled swapping.
Since each $j(\ge 2)$th iteration for quantum implementation of Gaussian elimination can takes $\alpha(m-j+1)(j-1)$ qubits as borrowed qubits, these qubits are used for the parallelization of controlled swaping.
Broadly speaking, for $\mathsf{Pivoting}$ with $m$ rows, there exist $\lceil\log_{2}m\rceil$ step for swapping and in $d$th step $\lfloor m / 2^{d} \rfloor$ rows are related for swapping.
For $d$th step, related swaps are parallelized with $\left\lceil \tfrac{g}{\lfloor m / 2^{d} \rfloor} \right\rceil$ borrowed qubits for each where $g$ is the total number of borrowed qubits.
The detailed algorithm is presented below.
\vspace{-15pt}
\begin{algorithm}[H]
	\small
	\caption{$\mathsf{ParallelPivoting}$}
	\begin{algorithmic}[1]
		\Require $\mathbf{A} \in \mathrm{GF}(p^{\ell})^{m \times n}$, $u$, $\mathbf{w} \in \{0,1\}^{m}$, $g$
		\Ensure $\mathbf{A}'' \in \mathrm{GF}(p^{\ell})^{m \times n}$, $\mathbf{w}'' \in \{0,1\}^{m}$
		\algrule
		\State $\mathbf{A}' \gets (\mathbf{A} \;|\; \mathbf{w})$
		\For{$d \in \{1, \ldots, \lceil \log_{2} m \rceil\}$}
			\State $\Delta \gets 2^{d-1}$
			\State $N_{\mathrm{pairs}} \gets \lfloor \tfrac{ m + 2^{d-1} - 1 } { \min( 2^{d}, m ) } \rfloor$
			\Comment{\# of row pairs at level $d$}
			\State $n_g \gets \lfloor g / N_{\mathrm{pairs}} \rfloor$
			\Comment{\# of borrowed qubits used for parallelization at level $d$}
			\For{$j \in \{1, \ldots, N_{\mathrm{pairs}}\}$}
				\Comment{$j$: row-pair iterator}
				\State $r_{1} \gets (j-1)2^{d} + 1$
				\State $r_{2} \gets r_{1} + \Delta$
				\If{$u = r_{2}$ \textbf{and} $r_{2} \le m$}
					\State $\mathbf{A}' \gets \mathsf{ParallelSwap}(\mathbf{A}', r_{1}, r_{2}, n_g)$
					\State $u \gets u - \Delta$
					\Comment{update target swap}
				\EndIf
			\EndFor
		\EndFor
		\State $(\mathbf{A}'' \;|\; \mathbf{w}'') \gets \mathbf{A}'$
		\State \textbf{return} $\mathbf{A}'', \mathbf{w}''$
	\end{algorithmic}
	\label{alg:prallelized_pivoting}
\end{algorithm}

In Algorithm~\ref{alg:prallelized_pivoting}, \textsf{ParallelSwap} is a subroutine for swapping two rows in parallel using borrowed qubits.
The underlying idea is introduced in Appendix~\ref{app:parallel-pivoting}, and the parallelization is developed accordingly.
Instead of giving a detailed account of the procedure, we illustrate its operation through a simple example shown in the figure.
\begin{figure}[htbp]
	\centering
	\includegraphics[width=0.8\textwidth]{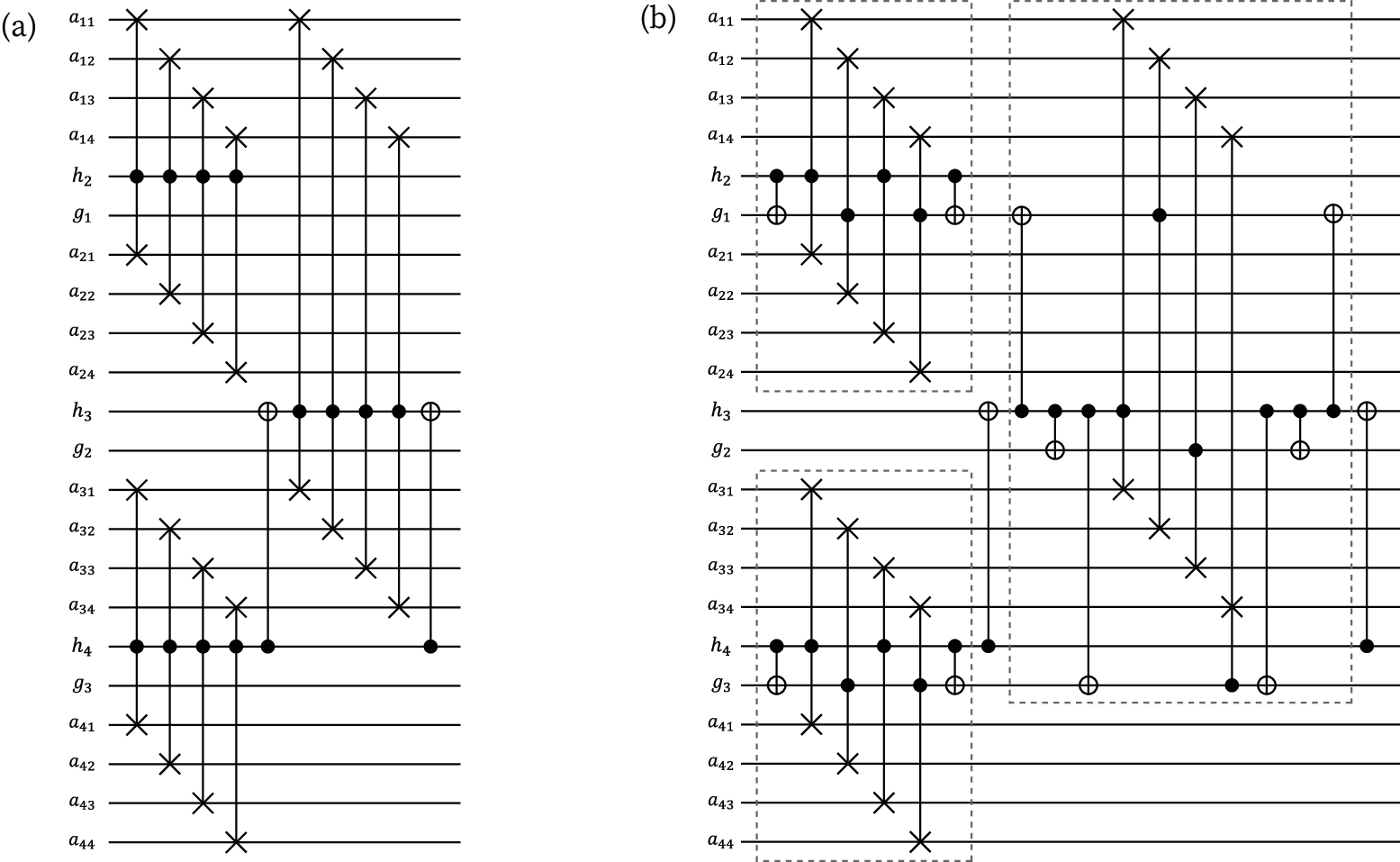}
	\caption{Example for quantum reversible implementation for $\mathsf{Pivoting}$ with $\mathbf{A} \in \mathrm{GF}(2)^{4 \times 4}$. (a) plain circuit for $\mathsf{Pivoting}$, (b) parallelized circuit for $\mathsf{Pivoting}$ with three borrowed qubits and gray dashed box presents parallelized swaps.}
\end{figure}

\subsubsection{Complexity}
The complexity of Algorithm~\ref{alg:prallelized_pivoting} is accounted for in Proposition~\ref{prop:parallel_pivot} in the main text.
Now we give a proof of Proposition~\ref{prop:parallel_pivot}.
\paragraph{Proof of Proposition~\ref{prop:parallel_pivot}.}
Let $m' := m-j'$ and $n' := n-j'$.
By Algorithm~\ref{alg:prallelized_pivoting}, \textsf{ParallelPivoting} proceeds through $\lceil \log_2 m' \rceil$ tournament levels, and there are $N_{d} = \left\lfloor \tfrac{m' + 2^{d-1} - 1}{\min( 2^d, m )} \right\rfloor$ candidate row pairs at level $d \in \{1,\dots,\lceil \log_2 m' \rceil\}$.
The tournament structure moves the target row upward by swaps, which takes place $m'-1$ times in total.
Each row consists of $n'$ field elements, and conditionally swapping one field element requires $\alpha$ controlled swaps (Fredkin gates).
In addition, the auxiliary register $w$ is also swapped for the bookkeeping.
Hence, one row swap is implemented by $\alpha n ' + 1$ Fredkin gates.
Since implementing a Fredkin gate entails one Toffoli gate, the total Toffoli count is bounded by $2(\alpha n' + 1)(m'-1)$,
where the factor 2 comes from the parallelized controlled swap construction using borrowed qubits, as described in Appendix~\ref{app:parallel-pivoting}.

We next analyze the depth.
At level $d$, the number of borrowed qubits available for each row pair is $n_g^{(d)} =
\left\lfloor
\tfrac{\alpha m'j'}{\left\lfloor m'/2^d \right\rfloor}
\right\rfloor$, because the total number of borrowed qubits is $\alpha m'j'$ and there are $N_{d}$ swap operations at level $d$.
Given that $n_g^{(d)}$ borrowed qubits are available, by Appendix~\ref{app:parallel-pivoting}, a row swap ($\alpha n'+1$ Fredkin) has Toffoli depth bounded by
\begin{align*}
	2\left\lceil
		\frac{\alpha n'+1}{n_g^{(d)}+1}
	\right\rceil 
=
	2\left\lceil
	\frac{\alpha n'+1}{\left\lfloor \tfrac{\alpha m'j'}{\lfloor m'/2^d\rfloor} \right\rfloor +1}
	\right\rceil
<
	\frac{n'}{j' \, 2^{\,d-1}} + 2 + \frac{2}{\alpha},
\end{align*}
where the last inequality is derived in a few steps which are omitted.
Summing over $d$, we finally have
\begin{align*}
	\sum_{d=1}^{\lceil \log_2 m' \rceil} 2\left\lceil
	\frac{\alpha n'+1}{\left\lfloor \tfrac{\alpha m'j'}{\lfloor m'/2^d\rfloor} \right\rfloor +1}
	\right\rceil
	<
	\sum_{d=1}^{\lceil \log_2 m' \rceil}
	\left(
	\frac{n'}{j'\,2^{\,d-1}} + 2 + \frac{2}{\alpha}
	\right)
	<
	\frac{2n'}{j'} + 3\lceil \log_2 m' \rceil .
\end{align*}
Substituting $m'=m-j'$ and $n'=n-j'$ back, we have Proposition~\ref{prop:parallel_pivot}.
$\qed$

\subsection{RowReduction}\label{sec:app-rowred}
\subsubsection{Algorithm}
The purpose of the $\mathsf{RowReduction}$ operation is to eliminate entries with respect to the first pivot row.
As the details of the $\mathsf{RowReduction}$ operation have already been presented in the main text, we include only the pseudocode below that works serially.
\begin{algorithm}[H]
	\small
	\caption{$\mathsf{SerialRowReduction}$}
	\begin{algorithmic}[1]
		\Require  $\mathbf{A} \in \text{GF}(p^{l})^{m \times n}$, $\mathbf{w} = (w_{2}, \ldots, w_{m}) \in \{0,1\}^{m-1}$
		\Ensure $\mathbf{A}' \in \text{GF}(p^{l})^{m \times n}$
		\Comment{pivot index in unary format}
		\algrule
		\State $\mathbf{A}' \gets \mathbf{A}$
		\For{$j \in \{1, \ldots, m\}$}
		\Comment{The first phase}
			\If {$w_{j} = 1$}
			\Comment{Suppose $w_{1} = 1$}
				\For{$i \in \{2, \ldots, n\}$}
					\State $a'_{ji} \gets a'_{ji} / a'_{j1}$
				\EndFor
			\EndIf
		\EndFor
		\For {$j \in \{2, \ldots, m\}$}
			\Comment{The second phase}
			\If {$w_{j} = 1$}
				\For {$i \in \{2, \ldots, n\}$}
					\State $a'_{ji} \gets a'_{ji} - a'_{1i}$
				\EndFor
			\EndIf
		\EndFor
		\State \textbf{return} $\mathbf{A}'$
	\end{algorithmic}
	\label{alg:row reduction}
\end{algorithm}

\subsubsection{In-Place Multiplication with Multiplicative Inverse}
At a high level, $\mathsf{RowReduction}$ consists of two phases.
In the first phase, each element of a row is multiplied by the multiplicative inverse of the first element of that row.
In the second phase, elimination is performed with respect to the pivot row.
While the second phase involves a straightforward arithmetic operation, the first phase requires further explanation.
In particular, the first phase is most naturally implemented as an in-place multiplication in order to minimize qubit usage.
However, quantum circuit implementations usually take out-of-place arithmetic operations as the basic primitives.
We therefore explain how the first phase operation can be realized using such out-of-place components with working qubits.

\begin{figure}[htbp]
	\centering
	\includegraphics[width=0.6\textwidth]{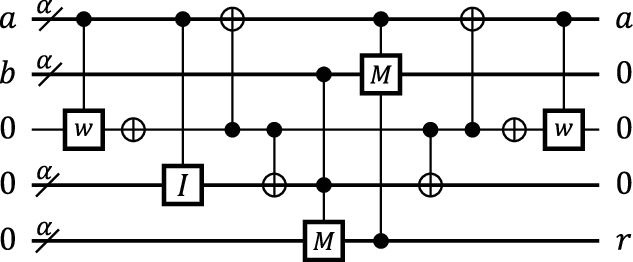}
	\caption{Gate-level realization of the transformation for computing $b(a^{-1}\oplus w\oplus 1)$.
		Here, the box $w$ computes the value of $w$ defined in Eq.~\eqref{eq:labeling-work}, the box $I$ denotes an out-of-place multiplicative-inverse operation, and the box $M$ denotes an out-of-place multiplication of two input values.
		The output register $r$ stores the final result $b(a^{-1}\oplus w\oplus 1)$, while all ancillary registers are restored to zero.}
	\label{fig:InPlaceInverseMut}
\end{figure}

In the first phase of $\mathsf{RowReduction}$, the target arithmetic operations can be accomplished by
\begin{align*}
	\ket{a}_{\alpha}\ket{b}_{\alpha}\ket{0}_{2\alpha+1}
	\;\longrightarrow\;
	\ket{a}_{\alpha}\ket{b\bigl(a^{-1}\oplus w\oplus 1\bigr)}_{\alpha}\ket{0}_{2\alpha+1},
\end{align*}
where $a,b \in \mathrm{GF}(p^{l})$, $\alpha = l\cdot \lceil \log_{2} p \rceil$, and $w$ is defined by Eq.~\eqref{eq:labeling-work}.
This construction relies on
\begin{align*}
	b\bigl(a^{-1}\oplus w\oplus 1\bigr)
	=
	\begin{cases}
	ba^{-1} & (a\neq 0)\\
	b & (a=0).
	\end{cases}
\end{align*}
Assuming that $0^{-1}=0$ in the circuit model and that out-of-place multiplication may accumulate into a nonzero target register, the operation can be implemented using two multiplicative-inverse operations, two multiplications, and four CNOT gates.
A detailed gate-level realization is illustrated in Fig.~\ref{fig:InPlaceInverseMut}.

We remark that, if $a=0$, then the first phase does not need to be applied to that row.
However, because the circuit cannot determine beforehand which row has a zero pivot, we adopt the convention that the multiplicative inverse module outputs $0$ on input $0$.
In this case, $a=0$, and the operation presented in Fig.~\ref{fig:InPlaceInverseMut} preserves the input value.
Therefore, the output is identical to the value prior to the multiplication by multiplicative inverse operation.


\subsubsection{Parallelization}
Based on the method for parallelizing arithmetic operations with borrowed qubits presented in Appendix~\ref{sec:app-ParBow}, the arithmetic operations in $\mathsf{RowReduction}$ can be parallelized.
In the $j$th iteration $j ~ ( \ge 2)$ of the quantum implementation of Gaussian elimination, there are $\alpha (m-j+1)(j-1) + \alpha (n-j)(j-1)$ qubits available as borrowed qubits, and these qubits can be exploited to parallelize $\mathsf{RowReduction}$.
More specifically, in the first phase of $\mathsf{RowReduction}$, the $n-1$ multiplications in each row are parallelized.
In the second phase, the $(m-1)(n-1)$ controlled additions are parallelized.
The detailed algorithm is presented below.
\begin{algorithm}[H]
	\small
	\caption{$\mathsf{ParallelRowReduction}$}
	\begin{algorithmic}[1]
		\Require  $\mathbf{A} \in \text{GF}(p^{l})^{m \times n}$, $\mathbf{w} = (w_{2}, \ldots, w_{m}) \in \{0,1\}^{m-1}$, $g_{1},g_{2},g_{3}$
		\Ensure $\mathbf{A}' \in \text{GF}(p^{l})^{m \times n}$
		\Comment{pivot index in unary format}
		\algrule
		\State $\mathbf{A}' \gets \mathbf{A}$
		\State $n_{g_{1}},\; n_{g_{2}},\; n_{g_{3}}  \gets \lfloor g_{1} / m \rfloor, \;\lfloor g_{2} / (m-1) \rfloor, \;\lfloor g_{3} / (n - 1) \rfloor$
		\For{$j \in \{1, \ldots, m\}$}
			\Comment{The first phase}
			\If {$w_{j} = 1$}
				\Comment{Suppose $w_{1} = 1$}
				\For{$i \in \{2, \ldots, n\}$}
					\State $a'_{ji} \gets \mathsf{ParallelMul}(a'_{ji}, a'_{j1}, n_{g_{1}})$
					\Comment{$a'_{ji} \gets a'_{ji} / a'_{1i}$}
				\EndFor
			\EndIf
		\EndFor
		\For {$j \in \{2, \ldots, m\}$}
			\Comment{The second phase}
			\If {$w_{j} = 1$}
				\For {$i \in \{2, \ldots, n\}$}
					\State $a'_{ji} \gets \mathsf{ParallelSub}(a'_{ji}, a'_{1i}, n_{g_{2}}, n_{g_{3}})$
					\Comment{$a'_{ji} \gets a'_{ji} - a'_{1i}$}
				\EndFor
			\EndIf
		\EndFor
	\end{algorithmic}
	\label{alg:row_reduction_par}
\end{algorithm}

\begin{figure}[htbp]
	\centering
	\includegraphics[width=0.9\textwidth]{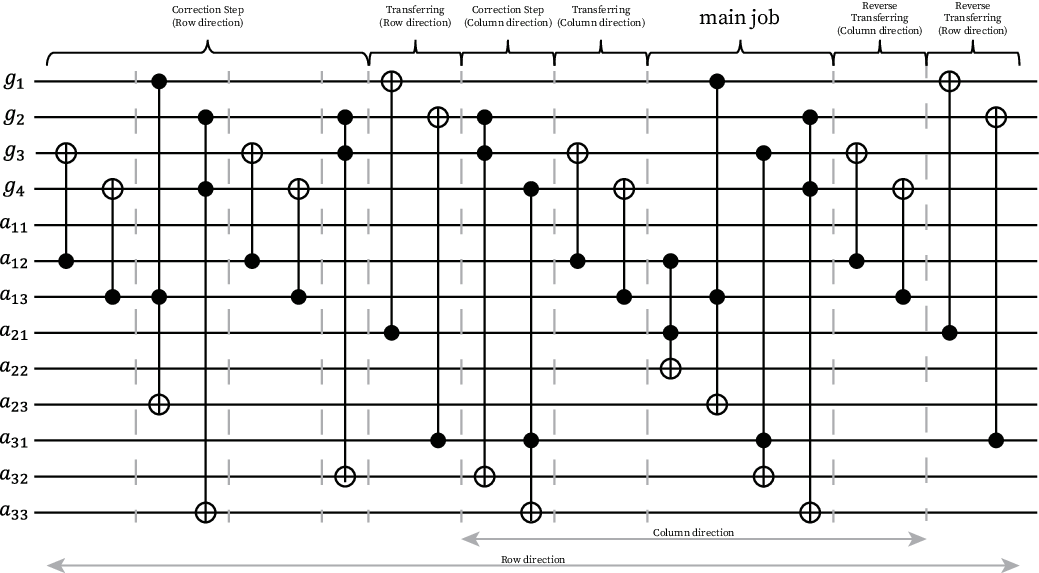}
	\caption{The example for parallelization of \textsf{ParallelRowReduction} with $\mathbf{A} \in \mathrm{GF}(2)^{3 \times 3}$ with 4 borrowed qubits.
	The gray dashed lines indicate the conceptual segments used for parallelization with borrowed qubits.}
	\label{fig:ExRowReducParal}
\end{figure}

In Algorithm~\ref{alg:row_reduction_par}, \textsf{ParallelMul} and \textsf{ParallelSub} are subroutines for each arithmetic operations for parallelization using borrowed qubits.
Moreover, the algorithm takes three input parameters, $g_{1}$, $g_{2}$, and $g_{3}$, representing the numbers of borrowed qubits used in the first phase, the row-direction parallelization of the second phase, and the column-direction parallelization of the second phase, respectively.
When $g_1$, $g_2$, and $g_3$ are chosen so that, accounting for the reuse of borrowed qubits, their effective total matches the number of borrowed qubits specified in Proposition~\ref{prop:row-reduction}, the algorithm achieves the depth bound stated there.
The justification can be found in the proof of Proposition~\ref{prop:row-reduction}.
Briefly, in the first phase, \textsf{ParallelMul} parallelizes the row-wise multiplications using $n_{g1}$ borrowed qubits, following the construction in Appendix~\ref{app:parallel-rowreduction}.
In the second phase, \textsf{ParallelSub} parallelizes the controlled additions in the row and column directions using $n_{g2}$ and $n_{g3}$ borrowed qubits, respectively, again following Appendix~\ref{app:parallel-rowreduction}.
	An example circuit over $\mathrm{GF}(2)$ is illustrated in Fig.~\ref{fig:ExRowReducParal}.
We remark that since the operations are done over $\mathrm{GF}(2)$, it does not need the first phase of \textsf{RowReduction}.
Therefore for $\mathbf{A} \in \mathrm{GF}(2)^{m \times n}$, it reads
\begin{align*}
	\left(\begin{matrix}
		a_{11} & a_{12} & \cdots & a_{1n}\\
		a_{21} & a_{22} & \cdots & a_{2n}\\
		\vdots & \vdots & \ddots & \vdots\\
		a_{m1} & a_{m2} & \cdots & a_{mn}
	\end{matrix}\right)
	&\longmapsto
	\left(\begin{matrix}
		a_{11} & a_{12} & \cdots & a_{1n}\\
		a_{21} & a_{22} \oplus a_{21}a_{12} & \cdots & a_{2n} \oplus a_{21}a_{1n}\\
		\vdots & \vdots & \ddots & \vdots\\
		a_{m1} & a_{m2} \oplus a_{m1}a_{12} & \cdots & a_{mn} \oplus a_{m1}a_{1n}
	\end{matrix}\right).
\end{align*}

\subsubsection{Complexity}
The complexity of Algorithm~\ref{alg:row_reduction_par} is accounted for in Proposition~\ref{prop:row-reduction} in the main text.
Before proving Proposition~\ref{prop:row-reduction}, it is helpful to consider an optimization method proposed in \cite{COG}.
Elimination on $\mathbf{A} \in \mathrm{GF}(p^{l})^{m \times n}$ requires $(m-1)(n-1)$ controlled subtractions.
Because these operations commute with one another, they can be scheduled efficiently by the coloring method and the resulting depth is $\max(m-1,n-1)$.

\paragraph{Proof of Proposition~\ref{prop:row-reduction}.}
Let $m' := m - j'$ and $n':=n-j'$.
We first consider the first phase (Phase~1 hereafter).
For each of the $m'$ rows, there are $n'-1$ entries (excluding the first column of $m' \times n'$ matrix) to be updated.
As described above, each multiplication by the multiplicative inverse is implemented using two out-of-place multiplications and two inverse operations.
Hence in the serial implementation, the multiplication count in Phase~1 is $2m'(n'-1)$.
Moreover, since the multiplicative inverse of the pivot element can be shared across in the same row, the number of inverse operations is $2m'$.
Be reminded that $\alpha m' j' + \alpha (n'-1) j'$ borrowed qubits are available.
For now, we use $\alpha m'j'$ qubits for parallelization in Phase 1.
Each row can use $\left\lfloor \tfrac{\alpha m'j'}{m'} \right\rfloor = \alpha j'$ borrowed qubits.
By the parallelized multiplication construction in Appendix~\ref{app:parallel-rowreduction}, the multiplication count is bounded by $4m'(n'-1)$, and the multiplication depth is bounded by $4\left\lceil \tfrac{n'-1}{j'+1} \right\rceil$.
About the inversion, each of $m'$ leading elements must be computed once and restored later, thus the multiplicative inverse count and depth are $2m'$ and $2$, respectively.
In addition, auxiliary addition (not controlled) is also introduced.
Using the construction in Appendix~\ref{app:distributing}, the auxiliary addition count and depth are bounded by $4m'j'$ and $4\lceil\log_{2}(j'+1)\rceil$, respectively.

Next we consider the second phase.
We first apply parallelization in the column direction (as introduced in Appendix~\ref{app:parallel-rowreduction}) by using $\alpha (n'-1)j'$ borrowed qubits.
For each column, $m'-1$ controlled additions take place.
Since the number of available borrowed qubits is sufficient to allocate $\left\lfloor \tfrac{\alpha (n'-1)j'}{n'-1} \right\rfloor = \alpha j'$ qubits to each column, Appendix~\ref{app:parallel-rowreduction} implies that the controlled additions in each column can be parallelized with its count and depth being bounded by $2(m'-1)$ and $2\left\lceil \tfrac{m'-1}{j'+1} \right\rceil$, respectively.
Therefore, over all $n'-1$ target columns, the total controlled addition count is bounded by $2(m'-1)(n'-1)$, while the depth is bounded by $2\max\left(\left\lceil \tfrac{m'-1}{j'+1} \right\rceil,n'-1\right)$.
In addition, auxiliary addition (not controlled) is also introduced.
Using the construction in Appendix~\ref{app:distributing}, the auxiliary addition count and depth are bounded by $4(n'-1)j'$ and $4 \lceil\log_{2}(j'+1)\rceil$, respectively.

We then apply the parallelization in the row direction by using $(m'-1)j'$ borrowed qubits.
Note that the borrowed qubits used in Phase 1 is reused here.
After the previous parallelization of column direction, each row involves at most $2(n'-1)$ controlled additions.
Since the number of available borrowed qubits is sufficient to allocate $\left\lfloor \tfrac{(m'-1)j'}{m'-1} \right\rfloor = j'$ qubits to each row, Appendix~\ref{app:parallel-rowreduction} implies that the controlled additions in each row can be parallelized with its count and depth being bounded by $4(n'-1)$ and $4\left\lceil \tfrac{n'-1}{j'+1} \right\rceil$, respectively.
Therefore, over all $m'-1$ target rows, the total controlled addition count is bounded by $4(m'-1)(n'-1)$, while the depth is bounded by $4\max\left( \left\lceil \tfrac{m'-1}{j'+1} \right\rceil, \left\lceil \tfrac{n'-1}{j'+1} \right\rceil\right)$.
In addition, auxiliary CNOT and controlled additive inverse operations are also introduced.
Using the construction in Appendix~\ref{app:distributing}, the CNOT count and depth are bounded by $4(m'-1) j'$ and $4\lceil\log_{2}(j'+1)\rceil$, respectively.
Similarly, the controlled additive inverse count and depth are bounded by $2(m'-1) (n'-1)$ and $2\left\lceil \tfrac{n'-1}{j'+1} \right\rceil$,	respectively.
Moreover, the auxiliary costs from the parallelization in column direction is doubled.
Therefore the auxiliary addition count is bounded by $8(n'-1) j'$
and the addition depth is bounded by $8\lceil\log_{2}(j'+1)\rceil$.

Collecting all bounds and substituting back $m'=m-j'$ and $n'=n-j'$, we have Proposition~\ref{prop:row-reduction}.
$\qed$

\bigskip
Below we give a proof of Remark~\ref{rem:row-reduction} which is the complexity result for row reduction in $j=1$.

\paragraph{Proof of Remark~\ref{rem:row-reduction}.}
We first consider Phase~1.
For each of the $m$ rows, there are $n-1$ entries that must be updated using the pivot element.
As in the proof of Proposition~\ref{prop:row-reduction}, each multiplication by the multiplicative inverse is implemented using two out-of-place multiplications and two inverse operations.
Hence, the total multiplication count is $2m(n-1)$, and the total multiplicative inverse count is $2m$.
Moreover, since the operations on different rows act on disjoint qubits, they can be executed in parallel across the rows.
Therefore, the multiplication depth is $2(n-1)$, and the multiplicative inverse depth is $2$.
Notice that since we do not have any borrowed qubit at $j=1$, the parallelization used in the proof of Proposition~\ref{prop:row-reduction} does not apply.

Next, we consider the second phase.
For each $i \in \{2,\dots,m\}$, the first row is conditionally subtracted from the $i$th row over the last $n-1$ columns.
Therefore, the total number of additions is $(m-1)(n-1)$.
Since these controlled subtractions commute with one another, they can be scheduled so that the resulting depth is $m-1=\max(m-1,\,n-1)$ when $m \ge n$.
$\qed$

\section{Explicit upper bounds}
\label{app:app-upper-bound}
\subsection{Upper bounds on the Toffoli cost of multi-controlled NOT gates}
\label{app:app-mcx-upper-bound}
In the complexity analysis of quantum Gaussian elimination, we use the asymptotic bounds for the cost of multi-controlled NOT gates.
Although asymptotic bounds are sufficient for the main results, explicit upper bounds are needed to derive concrete numerical estimates in Appendix~\ref{sec:app-ImplDetail}.
For this purpose, we summarize the recursive decomposition method introduced in~\cite{Claudon2024}.

Let $C(m)$ and $D(m)$ denote the Toffoli gate count and Toffoli depth of a $C^m \! X$ gate.
Using the decomposition of~\cite{Claudon2024}, we have
\begin{align*}
	C(m)
	&\le 2C(2p) + (4b-2)C(p) + 4C(b+c), \\
	D(m)
	&\le 2D(2p) + 6D(p),
\end{align*}
where $p = \lfloor \sqrt{m} \rfloor$, $b = \left\lfloor \tfrac{m-2p}{p} \right\rfloor + 1$, $c = \begin{cases} 0 & (m-2p) = 0\bmod p \\ 1 & \text{otherwise}\end{cases}$, and
%
\begin{align*}
	C(2) = D(2) = 1, \quad C(3) = D(3) = 4, \quad C(4) = D(4) = 10 .
\end{align*}

Since each recursion reduces the problem size from $m$ to approximately $\sqrt{m}$, the number of recursion levels is $O\big( \! \log_2 (\log_2 m)\big)$.
All things considered, the above recursion yields
\begin{align*}
	C(m) = O\!\left(m \log_2^2 m\right), \quad
	D(m) = O\!\left(\log_2^3 m\right).
\end{align*}
Hence, there exist constants $\kappa_C,\kappa_D > 0$ such that
\begin{align*}
	C(m) \le \kappa_C\, m \log_2^2 m, \quad
	D(m) \le \kappa_D\, \log_2^3 m.
\end{align*}
To find such constants, we numerically evaluate the above recursions for $m \in [2,10^7]$ and compare them with the reference functions $m\log_2^2 m$ and $\log_2^3 m$.
This yields the following explicit upper bounds over the range $[2,10^7]$:
\begin{align*}
	C(m) \le 2.185 \cdot m\log_2^2 m, \quad
	D(m) \le 3.121 \cdot \log_2^3 m.
\end{align*}


\subsection{Explicit upper bounds for Algorithm~\ref{alg:main-algorithm}}

We derive non-asymptotic upper bounds for the circuit cost.
Combining the results in Appendix~\ref{sec:app-ParBow}, Appendix~\ref{sec:app-ImplDetail}, we have Table~\ref{tab:complexity_par_bound} and Table~\ref{tab:complexity_par_bound2} each corresponding to non-asymptotic version of Table~\ref{tab:complexity_par} and Table~\ref{tab:complexity-comparison}, respectively.

\begin{table}[htbp]
	\renewcommand{\arraystretch}{1.3}
	\caption{
		Non-asymptotic bounds for Algorithm\;\ref{alg:main-algorithm} over $\mathrm{GF}(p^{l})$ where $p\neq 2$ and $\alpha = l \; \lceil\log_{2}p\rceil$.
	}
	\centering
	\begin{tabular}{
			>{\centering}m{1.8cm} |
			>{\centering}m{1.2cm} |
			>{\centering}m{9.0cm} }
		\hline
		\multirow{2}{*}{\makecell{\\ \scriptsize Toffoli}}
		& Count & \scriptsize{\makecell{
			$\tfrac{3\kappa_{C} n}{2} \log_{2}^{2}(\tfrac{m}{2}\!+\!2) \left( (2m\!-\!n\!+\!1) (\log_{2}m \!+\! 9) \!-\! 8 \right) + \tfrac{\alpha \kappa_{C} n}{2} (2m\!-\!n\!+\!1) \log_{2}^{2}\alpha $\\
			$+ (\alpha\!+\!1)(2mn\!-\!n^{2}\!-\!m\!-\!n\!+\!1) + \frac{n \alpha}{3}(3mn\!-\!n^{2}\!-\!3n\!+\!1)$
		}} \tabularnewline\cline{2-3}
		& Depth & \makecell{
			$6 \kappa_{D} n (\log_{2}(\tfrac{m}{2}\!+\!2))^{3}  \log_{2}(m\!+\!1) + \kappa_{D}n \log_{2}^{3}\alpha$\\
			$+ n((\alpha\!+\!2)\log_{2}m \!+\! 3\log_{2}m \!+\! 2\log_{2}n \!-\! 2) - 2\log_{2}m\!+\!2$
		} \tabularnewline\hline
		\multirow{2}{*}{\scriptsize Multiplication}
		& Count & $\tfrac{2}{3}(n\!-\!1)(3mn\!-\!n^{2}\!-\!3m\!+\!2n)$ \tabularnewline\cline{2-3}
		& Depth & $4n\log_{2}n \!-\! 2n \!-\! 2$ \tabularnewline\hline
		\multirow{2}{*}{\scriptsize \makecell{Multiplicative\\inverse}}
		& Count & $n(2m\!-\!n\!+\!1)$ \tabularnewline\cline{2-3}
		& Depth & $2n$ \tabularnewline\hline
		\multirow{2}{*}{\scriptsize \makecell{Controlled\\addition}}
		& Count & $\tfrac{1}{3}(n\!-\!1)(6mn\!-\!2n^{2}\!-\!9m\!-\!2n\!+\!9)$ \tabularnewline\cline{2-3}
		& Depth & $4m\log_{2}n\!-\!3m\!-\!1$ \tabularnewline\hline
		\multirow{2}{*}{\scriptsize \makecell{Additive\\inverse}}
		& Count & $\tfrac{1}{3}(n\!-\!1)(n\!-\!2)(3m\!-\!n\!-\!3)$ \tabularnewline\cline{2-3}
		& Depth & $2n\log_{2}n \!-\! 2n$ \tabularnewline\hline
	\end{tabular}
	\renewcommand{\arraystretch}{1.0}
	\label{tab:complexity_par_bound}
\end{table}

\begin{table}[htbp]
	\renewcommand{\arraystretch}{1.3}
	\caption{
		Non-asymptotic bounds for Algorithm\;\ref{alg:main-algorithm} over $\mathrm{GF}(2)$.
	}
	\centering
	\begin{tabular}{
			>{\centering}m{1.8cm} |
			>{\centering}m{1.2cm} |
			>{\centering}m{9.0cm} }
		\hline
		\multirow{2}{*}{\makecell{\\ Toffoli}}
		& Count & \makecell{
			$\tfrac{3\kappa_{C} n}{2} \log_{2}^{2}(\tfrac{m}{2}\!+\!2) \left( (2m\!-\!n\!+\!1) \! \log_{2}m \!+\! 18m \!-\! 9n \!+ \!1 \right)$\\
			$+ 3mn^{2} \!-\! n^{3} \!-\! 3mn \!-\! 2n^{2} \!+\! 2m \!+\! 3n \!-\! 2$
		} \tabularnewline\cline{2-3}
		& Depth & \makecell{
			$6 \kappa_{D} n (\log_{2}(\tfrac{m}{2}\!+\!2))^{3}  \log_{2}(m\!+\!1)$\\
			$+ 2(2m\!+\!n)\log_{2}n \!+\!2(2n\!-\!1)\log_{2}m \!-\! 3m \!-\! 2n \!+\! 1$
		} \tabularnewline\hline
	\end{tabular}
	\renewcommand{\arraystretch}{1.0}
	\label{tab:complexity_par_bound2}
\end{table}

\paragraph{Note on Labeling}
Non-asymptotic bound for \textsf{Labeling} is not messier than others.
Therefore, we leave a note on bounding it.

Let we introduce the following:
\begin{itemize}
	\item $C_{\rm lab}(m)$ : The number of Toffoli gates for $m$ bit \textsf{Labeling}
	\item $C_{\rm clab}(m)$ : The number of Toffoli gates for $m$ bit controlled \textsf{Labeling}
	\item $D_{\rm lab}(m)$ : The depth of Toffoli gates for $m$ bit \textsf{Labeling}
	\item $D_{\rm clab}(m)$ : The depth of Toffoli gates for $m$ bit cotrolled \textsf{Labeling}
\end{itemize}
Then we have
{\small
\begin{align}
	C_{\rm lab}(m) &= C_{\rm lab}(\lceil m/2\rceil -1) + C_{\rm clab}(\lfloor m/2 \rfloor) + C(\lceil m/2 \rceil) + 2 C(\lceil m/2 \rceil -1),\nn 
	C_{\rm clab}(m) &= C_{\rm clab}(\lceil m/2\rceil -1) + C_{\rm clab}(\lfloor m/2 \rfloor) + C(\lceil m/2 \rceil + 1) + 2 C(\lceil m/2 \rceil), \label{eq:noteLab2}\\
	D_{\rm lab}(m) &= D_{\rm clab}(\lfloor m/2 \rfloor) + D(\lceil m/2 \rceil) + 2 D(\lceil m/2 \rceil -1), \nn 
	D_{\rm clab}(m) &= D_{\rm clab}(\lfloor m/2 \rfloor) + D(\lceil m/2 \rceil+1) + 2 D(\lceil m/2 \rceil). \nonumber 
\end{align}
}

For $C_{\rm lab}(m)$, we have
\begin{align*}
	C_{\rm lab}(m)
	~<~ C_{\rm clab}(m)
	~<~ 2C_{\rm clab}(\lfloor m/2 \rfloor) + 3 C(\lceil m/2 \rceil + 1),
\end{align*}
where the first inequality is clear and the second inequality comes from loosening the right-hand side of Eq.~\eqref{eq:noteLab2}.
Note that
\begin{align*}
	\left\lfloor \frac{\lfloor m/2 \rfloor}{2} \right\rfloor \le \left\lfloor \frac{m}{4} \right\rfloor
	\quad\text{and}\quad
	\left\lceil \frac{\lfloor m/2 \rfloor}{2} \right\rceil +1 \le \left\lceil \frac{m}{4} \right\rceil + 1.
\end{align*}
Then we further loosen $C_{\rm clab}(\lfloor m/2 \rfloor)$ as
\begin{align*}
	C_{\rm clab}(\lfloor m/2 \rfloor)
	\quad<\quad 2C_{\rm clab}(\lfloor m/4 \rfloor) + 3 C(\lceil m/4 \rceil +1).
\end{align*}
By using above inequality, we can recursively bound $C_{\rm lab}(m)$ with $d$ steps as
\begin{align*}
	C_{\rm lab}(m)
	&< C_{\rm clab}(m)\\
	&< 2C_{\rm clab}(\lfloor m/2 \rfloor) + 3 C(\lceil m/2 \rceil + 1)\\
	&< 2^{2} C_{\rm clab} (\lfloor m /2^{2} \rfloor) + 3\left( 2 C(\lceil m/2^{2} \rceil + 1) + C(\lceil m/2 \rceil + 1) \right)\\
	&\qquad\vdots\\
	&< 2^{d} C_{\rm clab} (\lfloor m /2^{d} \rfloor) + 3 \textstyle\sum_{i=1}^{d} 2^{i-1} C(\lceil m/2^{i} \rceil + 1).
\end{align*}
Then, with $C_{\rm clab}(0) = 0$ and $C_{\rm clab}(1) = 1$ (see Fig.~\ref{fig:labeling}(c)), we have
\begin{align*}
	C_{\rm lab}(m)
	\;<\;
	3 \textstyle\sum_{i=1}^{\lceil\log_{2}m\rceil} 2^{i-1} \cdot C(\lceil m / 2^{i} \rceil + 1).
\end{align*}
Therefore, by Appedix~\ref{app:app-mcx-upper-bound}, we have
\begin{align*}
	C_{\rm lab}(m) < 3\kappa_{C} \; (\log_{2}(\tfrac{m}{2}+2))^{2} \left( \tfrac{m}{2} (\log_{2}m + 1) + (4m-2)  \right).
\end{align*}
Similarly, we have
\begin{align*}
	D_{\rm lab}(m) < 3 \; \kappa_{D} \; (\log_{2}(\tfrac{m}{2}+2))^{3} \; \log_{2}(m+1).
\end{align*}


\section{Correctness of Algorithm~\ref{alg:main-algorithm}}\label{app:CorrectnessOfMainAlgo}
Let $\mathbf{A}\in\mathrm{GF}(p^l)^{m\times n}$ satisfy $m\ge n$ and $\operatorname{rank}(\mathbf{A})=n$.
Algorithm~\ref{alg:main-algorithm} corresponds to classical algorithms that use the same pivot-selection rule and row-swap sequence as $\mathsf{SerialPivoting}$ (the pseudocode implementation of $\mathsf{Pivoting}$), together with a classical row-reduction procedure.
Here, $\mathsf{Labeling}$ returns the pivot index as an integer rather than in unary encoding, and this step can be carried out classically in a straightforward manner.

The tournament procedure in $\mathsf{ClassicPivoting}$ moves the selected row to the first row of the active submatrix.
At each level of the tournament, the selected row occupies the representative position of the merged block, and the final merged block is represented by its first row.
The assignment $u\gets r_1$ records the actual position of the selected row after each swap.

Next, $\mathsf{ClassicRowReduction}$ normalizes each row whose first entry is nonzero by dividing the row by that entry, and then subtracts the resulting pivot row from every remaining row whose first entry is $1$.
These elementary row operations preserve row equivalence, and their restriction to the active submatrix is valid because all previously processed columns vanish on the active rows.

At iteration $j$, the active submatrix has size $(m-j+1)\times(n-j+1)$ and full column rank $n-j+1$.
Hence, its first column contains a nonzero pivot, and elimination produces a block of the form $\begin{pmatrix}
1 & \mathbf{v}\\
0 & \mathbf{B}
\end{pmatrix}$,
where $\operatorname{rank}(\mathbf{B})=n-j$.
By induction, a pivot exists at every iteration, and the output $\mathbf{R}$ is row-equivalent to $\mathbf{A}$, has unit diagonal entries and zeros below the diagonal, and has zero rows at the bottom when $m>n$.
Consequently, $\mathbf{R}$ is in row-echelon form.

The following classical algorithm coincides with the corresponding steps of its quantum counterpart.

\begin{algorithm}[H]
	\small
	\caption{$\mathsf{RowEchelonForm}$}
	\begin{algorithmic}[1]
		\Require $\mathbf{A} \in \text{GF}(p^{l})^{m \times n}$, $m\ge n$, $\operatorname{rank}(\mathbf{A})=n$
		\Ensure A row echelon form $\mathbf{A}''$ row-equivalent to $\mathbf{A}$
		\algrule
		\State $\mathbf{A}' \gets \mathbf{A}$, $\mathbf{A}'' \gets \{0\}^{m\times n}$
		\For {$j \in \{1, \ldots, n\}$}
			\Comment{column iterator}
			\State $\mathmakebox[0pt][l]{u}\phantom{\mathbf{A}'} \gets \mathsf{Labeling}(\mathbf{A}')$
			\State $\mathbf{A}' \gets \mathsf{ClassicPivoting}(\mathbf{A}', u)$
			\State $\mathbf{A}' \gets \mathsf{ClassicRowReduction}(\mathbf{A}')$
			\State $\mathbf{A}''_{[j\,;\, j]} \gets \mathbf{A}'$
			\vspace{1mm}
			\State $\mathbf{A}' \gets \mathbf{A}'_{[2\, ;\, 2]}$
			\Comment{ignored if $j=n$}
		\EndFor
		\State \Return $\mathbf{A}''$
	\end{algorithmic}
	\label{alg:main-algorithm_ClassicVersion}
\end{algorithm}

\begin{algorithm}[H]
	\small
	\caption{$\mathsf{ClassicPivoting}$}
	\begin{algorithmic}[1]
		\Require $\mathbf{A} \in \text{GF}(p^{l})^{m \times n}$, $u=\min\{r:a_{r1}\ne0\}$
		\Ensure A row permutation $\mathbf{A}'$ of $\mathbf{A}$ with $a'_{11}\ne0$
		\algrule
		\State $\mathbf{A}' \gets \mathbf{A}$
		\For{$d \in \{1, \ldots, \lceil \log_{2}m \rceil \}$}
			\State $\Delta \gets 2^{d-1}$
			\For{$j \in \{1, \ldots, \lceil m/2^{d} \rceil\}$}
				\State $r_{2} \gets (m + 1 - \Delta) - (j-1)2^{d}$
				\State $r_{1} \gets \max(r_{2} - \Delta, 1)$
				\If{$u=r_{2}$ and $r_{1} \neq r_{2}$ and $r_{1}, r_{2} \ge 1$}
					\State $\mathbf{A}' \gets \mathsf{swap}(\mathbf{A}', r_{1}, r_{2})$
					\State $u \gets r_1$
				\EndIf
			\EndFor
		\EndFor
		\State \textbf{return} $\mathbf{A}'$
	\end{algorithmic}
	\label{alg:pivoting_ClassicVersion}
\end{algorithm}

\begin{algorithm}[H]
	\small
	\caption{$\mathsf{ClassicRowReduction}$}
	\begin{algorithmic}[1]
		\Require $\mathbf{A} \in \text{GF}(p^{l})^{m \times n}$, $a_{11}\ne0$
		\Ensure $\mathbf{A}'$ row-equivalent to $\mathbf{A}$ with first column $(1,0,\ldots,0)^{\mathsf{T}}$
		\algrule
		\State $\mathbf{A}' \gets \mathbf{A}$
		\For{$j \in \{1, \ldots, m\}$}
			\Comment{The first phase}
			\If {$a'_{j1} \neq 0$}
				\For{$i \in \{2, \ldots, n\}$}
					\State $a'_{ji} \gets a'_{ji} / a'_{j1}$
				\EndFor
			\State $a'_{j1} \gets 1$
			\Comment{after all divisions in this row}
			\EndIf
		\EndFor
		\For {$j \in \{2, \ldots, m\}$}
			\Comment{The second phase}
			\If {$a'_{j1} = 1$}
				\For {$i \in \{1, \ldots, n\}$}
					\State $a'_{ji} \gets a'_{ji} - a'_{1i}$
				\EndFor
			\EndIf
		\EndFor
		\State \textbf{return} $\mathbf{A}'$
	\end{algorithmic}
	\label{alg:row reduction_ClassicVersion}
\end{algorithm}

To establish the correctness of Algorithm~\ref{alg:main-algorithm}, it suffices to compare its action with that of the classical counterpart described above.
The classical algorithm uses the same pivot-selection rule and row-swap sequence as Algorithm~\ref{alg:main-algorithm}, and its row-reduction step consists only of elementary row operations.
As argued above, a valid pivot exists at every iteration, and the classical algorithm therefore produces a row-echelon form $\mathbf{R}$ that is row-equivalent to the input matrix $\mathbf{A}$.

The correspondence between the classical and quantum procedures can be established inductively over the iterations.
At the beginning of each iteration, their active submatrices coincide.
Since the same pivot is selected and the same row swaps are applied, this correspondence is preserved through the pivoting step.
Although $\mathsf{RowReduction}$ retains the entries in the current pivot column for reversibility, its action on the remaining columns agrees with that of $\mathsf{ClassicRowReduction}$.
In particular, both procedures normalize the rows having a nonzero pivot-column entry and subsequently eliminate the corresponding entries using the normalized pivot row.
Hence, they produce identical pivot-row entries to the right of the pivot and identical trailing active submatrices.
The subsequent uncomputation of the pivot information in Algorithm~\ref{alg:main-algorithm} affects neither of these parts, and thus preserves the correspondence for the next iteration.

Consequently, if $\mathbf{P}$ denotes the output of Algorithm~\ref{alg:main-algorithm}, then its entries relevant to the row-echelon structure coincide with those of $\mathbf{R}$; in particular,
\begin{align*}
    p_{ij} = r_{ij},
    \qquad 1\leq i<j\leq n.
\end{align*}
Since $\mathbf{R}$ is a row-echelon form of $\mathbf{A}$, the output $\mathbf{P}$ retains exactly the information required by Definition~\ref{def:PseudoRowEchelon}, while the remaining entries store the additional information required for reversibility.
Therefore, $\mathbf{P}$ is a pseudo row-echelon form of $\mathbf{A}$ in the sense of Definition~\ref{def:PseudoRowEchelon}.

\end{document}